\documentclass[prd,twocolumn,floatfix,amsmath,nofootinbib,amssymb,floatfix]{revtex4}
\usepackage{graphicx,color,dcolumn,booktabs,bm}
\usepackage{longtable,lscape}
\usepackage{pdfpages}
\usepackage{txfonts}
\usepackage{overpic}
\usepackage{amssymb}
\usepackage{makecell}
\usepackage{indentfirst}
\usepackage{feynmf}   
\usepackage{slashed}  
\usepackage{cases}
\usepackage{color}
\usepackage{multirow}
\usepackage{makecell}
\usepackage{threeparttable}
\usepackage{epstopdf}
\usepackage{enumerate}
\usepackage{diagbox}
\usepackage{graphicx,color,dcolumn,booktabs,bm}
\usepackage[colorlinks,
            citecolor=blue,
            anchorcolor=red,
            menucolor=red,
            linkcolor=red,
            filecolor=red,
            runcolor=red,
            urlcolor=blue,
            frenchlinks=red]{hyperref}
\usepackage{tikz}

\begin{document}

\title{Possible molecules of triple-heavy pentaquarks within the extended local hidden gauge formalism}

\author{Zhong-Yu Wang$^{1,2}$}
\email{zhongyuwang@foxmail.com}

\author{Chu-Wen Xiao$^{3,4,5}$}
\email{xiaochw@gxnu.edu.cn}

\author{Zhi-Feng Sun$^{1,2,6,7}$}
\email{sunzf@lzu.edu.cn}

\author{Xiang Liu$^{1,2,6,7}$}
\email{xiangliu@lzu.edu.cn}

\affiliation{
$^1$School of Physical Science and Technology, Lanzhou University, Lanzhou 730000, China \\
$^2$Lanzhou Center for Theoretical Physics, Key Laboratory of Theoretical Physics of Gansu Province, and Key Laboratory of Quantum Theory and Applications of the Ministry of Education, Lanzhou University, Lanzhou, 730000, China \\
$^3$Department of Physics, Guangxi Normal University, Guilin 541004, China \\
$^4$Guangxi Key Laboratory of Nuclear Physics and Technology,
Guangxi Normal University, Guilin 541004, China \\
$^5$School of Physics, Central South University, Changsha 410083, China \\
$^6$MoE Frontiers Science Center for Rare Isotopes, Lanzhou University, Lanzhou 730000, China \\
$^7$Research Center for Hadron and CSR Physics, Lanzhou University and Institute of Modern Physics of CAS, Lanzhou 730000, China
}

\date{\today}

\begin{abstract}

In this study, we explore the interactions between mesons and baryons in the open heavy sectors to identify potential triple-heavy molecular pentaquarks. We derive the meson-baryon interaction potentials using the vector meson exchange mechanism within the extended local hidden gauge formalism. The scattering amplitudes are computed by solving the coupled-channel Bethe-Salpeter equation, revealing several bound systems. By analyzing the poles of these amplitudes in the complex plane, we determine the masses and widths of these bound states. Additionally, we evaluate the couplings and compositeness of different channels within each bound system to assess their molecular characteristics. Our predictions include four $\Omega_{ccc}$-like states, four $\Omega_{bbb}$-like states, fourteen $\Omega_{bcc}$-like states, and ten $\Omega_{bbc}$-like states, which could be targets for future experimental investigations.

\end{abstract}
\maketitle

\section{Introduction}\label{sec:Introduction}

Over the past two decades, advancements in high-energy physics have led to the discovery of numerous new hadronic states. Many of these states do not conform to the traditional classifications of mesons, which consist of a quark and an antiquark, or baryons, which are made up of three quarks or antiquarks \cite{Gell-Mann:1964ewy,Zweig:1981pd}. Instead, these states may be candidates for exotic forms such as glueballs, hybrids, and multiquark states, which have garnered significant interest from both experimentalists and theorists \cite{Liu:2013waa,Hosaka:2016pey,Chen:2016qju,Richard:2016eis,Lebed:2016hpi,Olsen:2017bmm,Guo:2017jvc,Liu:2019zoy,Brambilla:2019esw,Meng:2022ozq,Chen:2022asf,Liu:2024uxn}. Understanding these states is crucial for advancing our knowledge of the non-perturbative behavior of strong interactions.

Among these reported states, a notable group includes hidden-charm hadrons, such as the charmonium-like $XYZ$ states and $P_c/P_{cs}$ pentaquark-like states, which have sparked extensive discussion regarding molecular multiquark states. For example, the $X(3872)$, the first experimentally observed charmonium-like $XYZ$ state, was reported by the Belle collaboration in 2003 through the decay $B^{\pm}\rightarrow K^{\pm}\pi^{+}\pi^{-}J/\psi$ \cite{Belle:2003nnu}. Its mass, close to the $D\bar{D}^{*}$ threshold, suggests it could be a $D\bar{D}^{*}$ molecular state, as proposed in several studies \cite{Swanson:2003tb,Wong:2003xk,Close:2003sg,Voloshin:2003nt,AlFiky:2005jd,Liu:2008fh,Thomas:2008ja,Liu:2009qhy,Lee:2009hy,Gamermann:2009uq,Braaten:2010mg,Baru:2013rta,Baru:2015nea,Song:2023pdq}.

In 2013, the BESIII and Belle collaborations reported the charge charmonium-like state $Z_c(3900)$, which is also near the $D\bar{D}^*$ threshold and can be interpreted as a molecular state \cite{BESIII:2013ris,Belle:2013yex,Guo:2013sya,Wang:2013daa,Aceti:2014uea}. Later, in 2015, the LHCb collaboration discovered the states $P_{c}(4380)^{+}$ and $P_{c}(4450)^{+}$ in the $J/\psi p$ invariant mass spectrum from the decay $\Lambda_{b}^{0}\rightarrow J/\psi pK^{-}$ \cite{LHCb:2015yax,LHCb:2016lve}. These states, initially predicted as molecular pentaquarks \cite{Wu:2010jy,Wu:2010vk,Wang:2011rga,Yang:2011wz,Wu:2012md,Xiao:2013yca,Li:2014gra,Karliner:2015ina}, have since received considerable theoretical attention, as discussed in reviews \cite{Hosaka:2016pey,Chen:2016qju,Olsen:2017bmm,Guo:2017jvc}.

In 2019, the LHCb collaboration further refined their observations, revealing that the previously identified $P_c$ states were actually three distinct states: $P_{c}(4312)^{+}$, $P_{c}(4440)^{+}$, and $P_{c}(4457)^{+}$, which provided strong support for the molecular interpretation \cite{LHCb:2019kea,Chen:2019asm,Chen:2019bip,Liu:2019tjn,He:2019ify,Xiao:2019mst,Guo:2019kdc,Xiao:2019aya}.

More recently, in 2021, the LHCb collaboration reported the hidden-charm pentaquark structure $P_{cs}(4459)^{0}$ in the decay $\Xi_{b}^{-}\rightarrow J/\psi \Lambda K^{-}$ \cite{LHCb:2020jpq}. Additionally, a new state, $P_{cs}(4338)^{0}$, containing a strange quark, was observed in the decay $B^{-}\rightarrow J/\psi \Lambda \bar{p}$ \cite{LHCb:2022ogu}. These states are considered strong candidates for $\bar{D}^{*}\Xi_{c}$ and $\bar{D}\Xi_{c}$ molecular states, respectively \cite{Xiao:2021rgp,Du:2021bgb,Zhu:2021lhd,Lu:2021irg,Zou:2021sha,Wang:2021itn,Wu:2021caw,Yan:2022wuz,Meng:2022wgl,Burns:2022uha,Ortega:2022uyu}. Thus, within the molecular framework and flavor symmetry, it is natural to assign the $P_{c}(4312)$ and $P_{c}(4440)$, $P_{c}(4457)$ as $\bar{D}\Sigma_c$ and $\bar{D}^{*}\Sigma_c$ states, respectively, and the $P_{cs}(4338)$ and $P_{cs}(4459)$ as $\bar{D}\Xi_{c}$ and $\bar{D}^{*}\Xi_{c}$ molecules, respectively, replacing the light quark with a strange quark, $q \to s$.

\begin{figure}[htbp]
\centering
\includegraphics[width=0.48\textwidth]{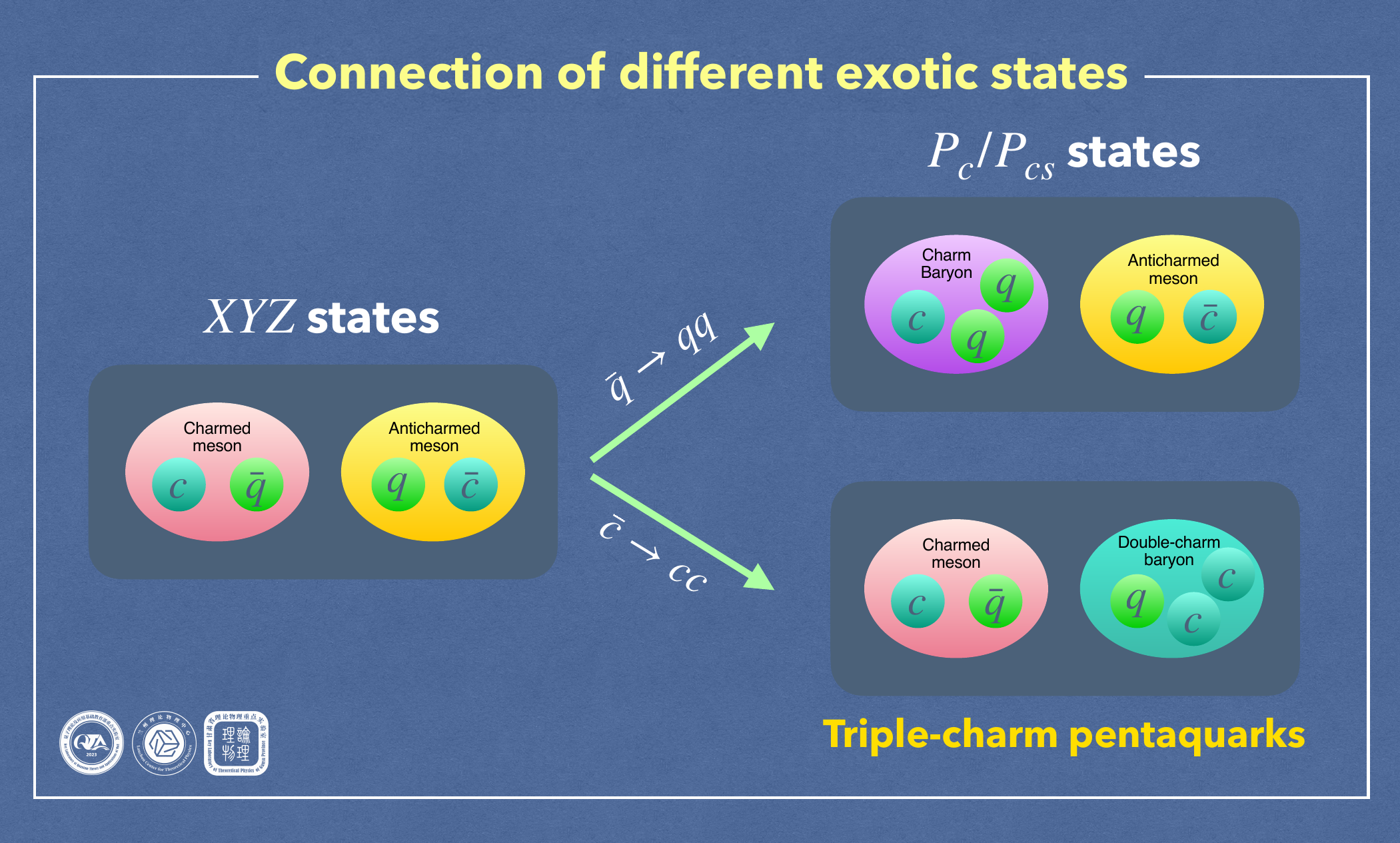} 
\caption{Connection of different heavy flavor hadronic molecular states, just with the substitutions $\bar{q}\to qq$ and $\bar{c}\to cc$ in the molecular picture.}
\label{fig:Figures}
\end{figure} 

From the molecular perspective, the hidden-charm pentaquark states $P_{c}$ and $P_{cs}$ can be seen as a straightforward substitution of $\bar{q} \rightarrow qq$ in the $XYZ$ states, as illustrated in Fig. \ref{fig:Figures}. Similarly, by substituting $\bar{c} \rightarrow cc$ in the $XYZ$ states, we obtain triple-charm pentaquark states through the interaction between a charmed meson and a double-charm baryon, which is of significant interest.

In 2013, Ref. \cite{Guo:2013xga} predicted the existence of exotic triple-heavy pentaquarks based on heavy antiquark–diquark symmetry, assuming that the $X(3872)$ and $Z_{b}(10610/10650)$ were molecular states. Subsequently, Refs. \cite{Chen:2017jjn,Wang:2019aoc} explored this further, predicting isoscalar triple-charm molecular-type pentaquark candidates such as $D^{(*)}\Xi_{cc}$, $D_{1}\Xi_{cc}$, and $D_{2}^{*}\Xi_{cc}$ using the one-boson-exchange model. These predictions offer valuable insights for future experiments.

Building on these predictions and the motivation from the findings of the $P_c$ and $P_{cs}$ states, we investigate triple-heavy pentaquark systems in the present work. Specifically, we consider pentaquark systems with quark contents $cccq\bar{q}$, $bbbq\bar{q}$, $bccq\bar{q}$, and $bbcq\bar{q}$, which we refer to as $\Omega_{ccc}$-like, $\Omega_{bbb}$-like, $\Omega_{bcc}$-like, and $\Omega_{bbc}$-like molecular states. For the meson-baryon interaction, we use a coupled-channel approach with transition potentials derived from the extended local hidden gauge formalism. Previous studies \cite{Wang:2022aga,Wang:2023mdj,Song:2024yli} have predicted many molecular pentaquark states with double heavy quarks using this formalism.

Our paper is constructed as follows.  
In Sec. \ref{sec:Formalism}, we introduce the $S$-wave interactions of the meson-baryon derived from the extended local hidden gauge formalism.
Then, in Sec. \ref{sec:Results}, we give the results of the triple-heavy pentaquark candidates by solving the coupled channel Bethe-Salpeter equation.
Finally, a short summary is shown in Sec. \ref{sec:Summary}.

\section{Formalism}\label{sec:Formalism}

In the present work, we search for the triple-heavy pentaquarks, including the $\Omega_{ccc}$-like states with the quark contents $cccq\bar{q}$ ($q=u,d,s$), the $\Omega_{bbb}$-like states with $bbbq\bar{q}$, the $\Omega_{bcc}$-like states with $bccq\bar{q}$, and the $\Omega_{bbc}$-like states with $bbcq\bar{q}$. 
Thus, the corresponding meson-baryon coupled channels involved for these quark systems are listed in Table~\ref{tab:coupledchannelsbcc}, where $P$ and $V$ denote the pseudoscalar and vector mesons, respectively, while $B(\frac{1}{2}^{+})$ and $B(\frac{3}{2}^{+})$ stand for the baryon ground states with $J^{P}=\frac{1}{2}^{+}$ and $J^{P}=\frac{3}{2}^{+}$, respectively. 
Additionally, the corresponding thresholds of these coupled channels are also given, where some of the masses are taken from the Particle Data Group (PDG) \cite{ParticleDataGroup:2022pth} and the others from the theoretical results with the constituent quark model \cite{Roncaglia:1995az,Roberts:2007ni,Lu:2017meb,Weng:2018mmf,Yang:2019lsg}.
There are four blocks of the coupled channels interactions, denoting as $PB(\frac{1}{2}^{+})$, $PB(\frac{3}{2}^{+})$, $VB(\frac{1}{2}^{+})$, and $VB(\frac{3}{2}^{+})$ as shown in Table~\ref{tab:coupledchannelsbcc}.

\begin{table*}[htbp]
\centering
\renewcommand\tabcolsep{4mm}
\renewcommand{\arraystretch}{1.50}
\caption{The coupled channels and the corresponding thresholds (in MeV) involved in the $\Omega_{ccc}$-like, $\Omega_{bbb}$-like, $\Omega_{bcc}$-like, and $\Omega_{bbc}$-like sectors.}
\begin{tabular*}{178mm}{c|cccccccc}
\toprule[1.00pt]
\toprule[1.00pt]
&\multicolumn{4}{c}{\mbox{$\Omega_{ccc}$-like sector}}&\multicolumn{4}{c}{\mbox{$\Omega_{bbb}$-like sector}}\\
\hline
\multirow{2}{*}{$PB(\frac{1}{2}^{+})$}&$D\Xi_{cc}$&$D_{s}\Omega_{cc}$&&&$\bar{B}\Xi_{bb}$&$\bar{B}_{s}\Omega_{bb}$&&\\
&$5489.25$&$5683.35$&&&$15619.50$&$15596.92$&&\\
\multirow{2}{*}{$PB(\frac{3}{2}^{+})$}&$\eta\Omega_{ccc}$&$\eta^{'}\Omega_{ccc}$&$D\Xi_{cc}^{*}$&$D_{s}\Omega_{cc}^{*}$&$\eta\Omega_{bbb}$&$\eta^{'}\Omega_{bbb}$&$\bar{B}\Xi_{bb}^{*}$&$\bar{B}_{s}\Omega_{bb}^{*}$\\
&$5345.86$&$5755.78$&$5542.25$&$5740.35$&$14943.86$&$15353.78$&$15649.50$&$15624.92$\\
\multirow{2}{*}{$VB(\frac{1}{2}^{+})$}&$D^{*}\Xi_{cc}$&$D_{s}^{*}\Omega_{cc}$&&&$\bar{B}^{*}\Xi_{bb}$&$\bar{B}_{s}^{*}\Omega_{bb}$&&\\
&$5630.56$&$5827.20$&&&$15664.71$&$15645.40$&&\\
\multirow{2}{*}{$VB(\frac{3}{2}^{+})$}&$\omega\Omega_{ccc}$&$\phi\Omega_{ccc}$&$D^{*}\Xi_{cc}^{*}$&$D_{s}^{*}\Omega_{cc}^{*}$&$\omega\Omega_{bbb}$&$\phi\Omega_{bbb}$&$\bar{B}^{*}\Xi_{bb}^{*}$&$\bar{B}_{s}^{*}\Omega_{bb}^{*}$\\
&$5580.66$&$5817.46$&$5683.56$&$5884.20$&$15178.66$&$15415.46$&$15694.71$&$15673.40$\\\hline
&\multicolumn{8}{c}{\mbox{$\Omega_{bcc}$-like sector}}\\
\hline
\multirow{2}{*}{$PB(\frac{1}{2}^{+})$}&$\eta\Omega_{bcc}$&$\eta^{'}\Omega_{bcc}$&$D\Xi_{bc}$&$D\Xi_{bc}^{'}$&$\bar{B}\Xi_{cc}$&$D_{s}\Omega_{bc}$&$D_{s}\Omega_{bc}^{'}$&$\bar{B}_{s}\Omega_{cc}$\\
&$8551.86$&$8961.78$&$8789.25$&$8815.25$&$8901.50$&$8979.35$&$9015.35$&$9081.92$\\
\multirow{2}{*}{$PB(\frac{3}{2}^{+})$}&$\eta\Omega_{bcc}^{*}$&$\eta^{'}\Omega_{bcc}^{*}$&$D\Xi_{bc}^{*}$&$\bar{B}\Xi_{cc}^{*}$&$D_{s}\Omega_{bc}^{*}$&$\bar{B}_{s}\Omega_{cc}^{*}$\\
&$8570.86$&$8980.78$&$8840.25$&$8954.50$&$9034.35$&$9138.92$\\
\multirow{2}{*}{$VB(\frac{1}{2}^{+})$}&$\omega\Omega_{bcc}$&$\phi\Omega_{bcc}$&$D^{*}\Xi_{bc}$&$D^{*}\Xi_{bc}^{'}$&$\bar{B}^{*}\Xi_{cc}$&$D_{s}^{*}\Omega_{bc}$&$D_{s}^{*}\Omega_{bc}^{'}$&$\bar{B}_{s}^{*}\Omega_{cc}$\\
&$8786.66$&$9023.46$&$8930.56$&$8956.56$&$8946.71$&$9123.20$&$9159.20$&$9130.40$\\
\multirow{2}{*}{$VB(\frac{3}{2}^{+})$}&$\omega\Omega_{bcc}^{*}$&$\phi\Omega_{bcc}^{*}$&$D^{*}\Xi_{bc}^{*}$&$\bar{B}^{*}\Xi_{cc}^{*}$&$D_{s}^{*}\Omega_{bc}^{*}$&$\bar{B}_{s}^{*}\Omega_{cc}^{*}$\\
&$8805.66$&$9042.46$&$8981.56$&$8999.71$&$9178.20$&$9187.40$\\
\hline
&\multicolumn{8}{c}{\mbox{$\Omega_{bbc}$-like sector}}\\
\hline
\multirow{2}{*}{$PB(\frac{1}{2}^{+})$}&$\eta\Omega_{bbc}$&$\eta^{'}\Omega_{bbc}$&$D\Xi_{bb}$&$\bar{B}\Xi_{bc}$&$\bar{B}\Xi_{bc}^{'}$&$D_{s}\Omega_{bb}$&$\bar{B}_{s}\Omega_{bc}$&$\bar{B}_{s}\Omega_{bc}^{'}$\\
&$11747.86$&$12157.78$&$12207.25$&$12201.50$&$12227.50$&$12198.35$&$12377.92$&$12413.92$\\
\multirow{2}{*}{$PB(\frac{3}{2}^{+})$}&$\eta\Omega_{bbc}^{*}$&$\eta^{'}\Omega_{bbc}^{*}$&$D\Xi_{bb}^{*}$&$\bar{B}\Xi_{bc}^{*}$&$D_{s}\Omega_{bb}^{*}$&$\bar{B}_{s}\Omega_{bc}^{*}$\\
&$11768.86$&$12178.78$&$12237.25$&$12252.50$&$12226.35$&$12432.92$\\
\multirow{2}{*}{$VB(\frac{1}{2}^{+})$}&$\omega\Omega_{bbc}$&$\phi\Omega_{bbc}$&$D^{*}\Xi_{bb}$&$\bar{B}^{*}\Xi_{bc}$&$\bar{B}^{*}\Xi_{bc}^{'}$&$D_{s}^{*}\Omega_{bb}$&$\bar{B}_{s}^{*}\Omega_{bc}$&$\bar{B}_{s}^{*}\Omega_{bc}^{'}$\\
&$11982.66$&$12219.46$&$12348.56$&$12246.71$&$12272.71$&$12342.20$&$12426.40$&$12462.40$\\
\multirow{2}{*}{$VB(\frac{3}{2}^{+})$}&$\omega\Omega_{bbc}^{*}$&$\phi\Omega_{bbc}^{*}$&$D^{*}\Xi_{bb}^{*}$&$\bar{B}^{*}\Xi_{bc}^{*}$&$D_{s}^{*}\Omega_{bb}^{*}$&$\bar{B}_{s}^{*}\Omega_{bc}^{*}$\\
&$12003.66$&$12240.46$&$12378.55$&$12297.71$&$12370.20$&$12481.40$\\
\bottomrule[1.00pt]
\bottomrule[1.00pt]
\end{tabular*}
\label{tab:coupledchannelsbcc}
\end{table*}

\begin{figure}[htbp]
\centering
\includegraphics[width=0.8\linewidth,trim=150 580 250 120,clip]{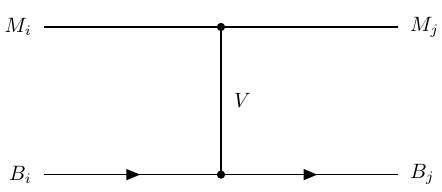} 
\caption{The $M_{i}B_{i}\rightarrow M_{j}B_{j}$ scattering via the vector meson exchange. $M_{i}(M_{j})$ and $B_{i}(B_{j})$ are the initial (final) meson and baryon, respectively.}
\label{fig:MB}
\end{figure} 

For the $M_{i}B_{i}\rightarrow M_{j}B_{j}$ transiton as shown in Fig. \ref{fig:MB}, we only consider the contribution from the vector meson exchange at the tree level, where the contribution from the pseudoscalar meson exchange is ignored as done in Refs. \cite{Xiao:2013yca,Debastiani:2017ewu}.
Thus, for our cases, there are three typical interaction vertices involved, such as the ones of $VPP$, $VVV$, and $VBB$. Using the extended local hidden gauge approach \cite{Liang:2017ejq,Dias:2018qhp,Yu:2018yxl,Yu:2019yfr,Dias:2019klk,Liang:2020dxr}, the Lagrangians of the vertices $VPP$ and $VVV$ can be expressed as
\begin{equation}
\begin{aligned} 
\mathcal{L}_{V P P}=-i g\left\langle\left[P, \partial_\mu P\right] V^\mu\right\rangle,
\end{aligned}
\label{eq:LVVP}
\end{equation}
\begin{equation}
\begin{aligned} 
\mathcal{L}_{V V V}=i g\left\langle\left(V^\mu \partial_\nu V_\mu-\partial_\nu V^\mu V_\mu\right) V^\nu\right\rangle,
\end{aligned}
\label{eq:LVVV}
\end{equation}
respectively, where $g=M_{V}/(2f_{\pi})$ is the coupling constant, $M_{V}$ the mass of the exchanged light vector meson and $f_{\pi}=93$ MeV the pion decay constant.
The notation $\left\langle \cdots \right\rangle$ in the expressions denotes the trace of a matrix. Additionally, $P$ and $V^{\mu}$ are pseudoscalar and vector meson fields matrices, respectively, which can be extended from SU(3) to SU(5), i.e.,
\begin{equation}
\begin{aligned}
P=\left(\begin{array}{ccccc}
\frac{\eta}{\sqrt{3}}+\frac{\eta^{\prime}}{\sqrt{6}}+\frac{\pi^0}{\sqrt{2}} & \pi^{+} & K^{+} & \bar{D}^0 & B^{+} \\
\pi^{-} & \frac{\eta}{\sqrt{3}}+\frac{\eta^{\prime}}{\sqrt{6}}-\frac{\pi^0}{\sqrt{2}} & K^0 & D^{-} & B^0 \\
K^{-} & \bar{K}^0 & -\frac{\eta}{\sqrt{3}}+\sqrt{\frac{2}{3}}\eta^{\prime} & D_s^{-} & B_s^0 \\
D^0 & D^{+} & D_s^{+} & \eta_c & B_c^{+} \\
B^{-} & \bar{B}^0 & \bar{B}_s^0 & B_c^{-} & \eta_b
\end{array}\right),
\end{aligned}
\label{eq:P}
\end{equation}
\begin{equation}
\begin{aligned}
V^{\mu}=\left(\begin{array}{ccccc}
\frac{\omega+\rho^0}{\sqrt{2}} & \rho^{+} & K^{*+} & \bar{D}^{* 0} & B^{*+} \\
\rho^{-} & \frac{\omega-\rho^0}{\sqrt{2}} & K^{* 0} & D^{*-} & B^{* 0} \\
K^{*-} & \bar{K}^{* 0} & \phi & D_s^{*-} & B_s^{* 0} \\
D^{* 0} & D^{*+} & D_s^{*+} & J / \psi & B_c^{*+} \\
B^{*-} & \bar{B}^{* 0} & \bar{B}_s^{* 0} & B_c^{*-} & \Upsilon
\end{array}\right)^{\mu}.
\end{aligned}
\label{eq:V}
\end{equation}
Note that Eqs.~\eqref{eq:P} and \eqref{eq:V} are the matrices of $q\bar{q}$ written in terms of mesons and that the results emerging from Eqs.~\eqref{eq:LVVP} and \eqref{eq:LVVV}, can be obtained simply using the $q\bar{q}$ content of the mesons without invoking SU(5) symmetry as shown in \cite{Sakai:2017avl}.
Therefore, the vertices $VPP$ and $VVV$ can be obtained from the effective Lagrangians of Eqs.~\eqref{eq:LVVP} and \eqref{eq:LVVV}.

However, for the vertex of $VBB$, it is difficult to directly extend the corresponding baryon fields matrix from the flavor SU(3) notation to the one of SU(5) with the charm and beauty baryon fields.
Therefore, instead of the evaluation from the effective Lagrangian, we use the scheme of Ref. \cite{Debastiani:2017ewu} to calculate the $VBB$ vertex from the flavor and spin wave functions of baryons used in \cite{Capstick:1986ter,Roberts:2007ni} in which the heavy quarks are singled out and the symmetry of identical particles is imposed on the light quarks. 
The baryons involved in our work, along with their explicit flavor and spin wave functions, are listed in Table \ref{tab:Wavefunctions},
where $\chi_{MS}$ indicates that the spin wave function is mixed symmetric, and $\chi_{MA}$ represents the mixed antisymmetric spin wave function. 
For the $J^{P}=\frac{1}{2}^{+}$ ground baryon states in the case of $S_{z}=+\frac{1}{2}$, the corresponding spin wave functions are given by
\begin{table}[htbp]
\centering
\renewcommand\tabcolsep{2mm}
\renewcommand{\arraystretch}{1.50}
\caption{The wave functions for baryon states. Here, $\chi_{MS}$ is mixed symmetric, $\chi_{MA}$ is  mixed antisymmetric, and $\chi_{S}$ is fully symmetric.}
\begin{tabular*}{86mm}{@{\extracolsep{\fill}}lccc}
\toprule[1.00pt]
\toprule[1.00pt]
States&$I(J^{P})$&Flavor& Spin\\
\hline
$\Xi_{cc}^{++}$&$\frac{1}{2}(\frac{1}{2}^{+})$&$ccu$&$\chi_{MS}(12)$\\
$\Xi_{cc}^{+}$&$\frac{1}{2}(\frac{1}{2}^{+})$&$ccd$&$\chi_{MS}(12)$\\
$\Omega_{cc}^{+}$&$0(\frac{1}{2}^{+})$&$ccs$&$\chi_{MS}(12)$\\
$\Xi_{bb}^{0}$&$\frac{1}{2}(\frac{1}{2}^{+})$&$bbu$&$\chi_{MS}(12)$\\
$\Xi_{bb}^{-}$&$\frac{1}{2}(\frac{1}{2}^{+})$&$bbd$&$\chi_{MS}(12)$\\
$\Omega_{bb}^{-}$&$0(\frac{1}{2}^{+})$&$bbs$&$\chi_{MS}(12)$\\
$\Xi_{bc}^{+}$&$\frac{1}{2}(\frac{1}{2}^{+})$&$\frac{1}{\sqrt{2}}b(cu-uc)$&$\chi_{MA}(23)$\\
$\Xi_{bc}^{0}$&$\frac{1}{2}(\frac{1}{2}^{+})$&$\frac{1}{\sqrt{2}}b(cd-dc)$&$\chi_{MA}(23)$\\
$\Xi_{bc}^{'+}$&$\frac{1}{2}(\frac{1}{2}^{+})$&$\frac{1}{\sqrt{2}}b(cu+uc)$&$\chi_{MS}(23)$\\
$\Xi_{bc}^{'0}$&$\frac{1}{2}(\frac{1}{2}^{+})$&$\frac{1}{\sqrt{2}}b(cd+dc)$&$\chi_{MS}(23)$\\
$\Omega_{bc}^{0}$&$0(\frac{1}{2}^{+})$&$\frac{1}{\sqrt{2}}b(cs-sc)$&$\chi_{MA}(23)$\\
$\Omega_{bc}^{'0}$&$0(\frac{1}{2}^{+})$&$\frac{1}{\sqrt{2}}b(cs+sc)$&$\chi_{MS}(23)$\\
$\Omega_{bcc}^{+}$&$0(\frac{1}{2}^{+})$&$bcc$&$\chi_{MS}(23)$\\
$\Omega_{bbc}^{0}$&$0(\frac{1}{2}^{+})$&$bbc$&$\chi_{MS}(12)$\\
$\Xi_{cc}^{*++}$&$\frac{1}{2}(\frac{3}{2}^{+})$&$ccu$&$\chi_{S}$\\
$\Xi_{cc}^{*+}$&$\frac{1}{2}(\frac{3}{2}^{+})$&$ccd$&$\chi_{S}$\\
$\Omega_{cc}^{*+}$&$0(\frac{3}{2}^{+})$&$ccs$&$\chi_{S}$\\
$\Xi_{bb}^{*0}$&$\frac{1}{2}(\frac{3}{2}^{+})$&$bbu$&$\chi_{S}$\\
$\Xi_{bb}^{*-}$&$\frac{1}{2}(\frac{3}{2}^{+})$&$bbd$&$\chi_{S}$\\
$\Omega_{bb}^{*-}$&$0(\frac{3}{2}^{+})$&$bbs$&$\chi_{S}$\\
$\Xi_{bc}^{*+}$&$\frac{1}{2}(\frac{3}{2}^{+})$&$\frac{1}{\sqrt{2}}b(cu+uc)$&$\chi_{S}$\\
$\Xi_{bc}^{*0}$&$\frac{1}{2}(\frac{3}{2}^{+})$&$\frac{1}{\sqrt{2}}b(cd+dc)$&$\chi_{S}$\\
$\Omega_{bc}^{*0}$&$0(\frac{3}{2}^{+})$&$\frac{1}{\sqrt{2}}b(cs+sc)$&$\chi_{S}$\\
$\Omega_{bcc}^{*+}$&$0(\frac{3}{2}^{+})$&$bcc$&$\chi_{S}$\\
$\Omega_{bbc}^{*0}$&$0(\frac{3}{2}^{+})$&$bbc$&$\chi_{S}$\\
$\Omega_{ccc}^{++}$&$0(\frac{3}{2}^{+})$&$ccc$&$\chi_{S}$\\
$\Omega_{bbb}^{-}$&$0(\frac{3}{2}^{+})$&$bbb$&$\chi_{S}$\\
\bottomrule[1.00pt]
\bottomrule[1.00pt]
\end{tabular*}
\label{tab:Wavefunctions}
\end{table}
\begin{equation}
\begin{aligned}
\chi_{MS}(12)=\frac{1}{\sqrt{6}}(\uparrow\downarrow\uparrow+\downarrow\uparrow\uparrow-2\uparrow\uparrow\downarrow),
\end{aligned}
\label{eq:chiMS12}
\end{equation}
\begin{equation}
\begin{aligned}
\chi_{MS}(23)=\frac{1}{\sqrt{6}}(\uparrow\downarrow\uparrow+\uparrow\uparrow\downarrow-2\downarrow\uparrow\uparrow),
\end{aligned}
\label{eq:chiMS23}
\end{equation}
\begin{equation}
\begin{aligned}
\chi_{MA}(23)=\frac{1}{\sqrt{2}}(\uparrow\uparrow\downarrow-\uparrow\downarrow\uparrow).
\end{aligned}
\label{eq:chiMA23}
\end{equation}
And $\chi_{S}$ is the fully symmetric spin wave function. In the case of $S_{z}=+\frac{3}{2}$ for the $J^{P}=\frac{3}{2}^{+}$ ground baryons, we have
\begin{equation}
\begin{aligned}
\chi_{S}=\uparrow\uparrow\uparrow.
\end{aligned}
\label{eq:chiS}
\end{equation}
The overlap of these spin wave functions are given by
\begin{equation}
\begin{aligned}
\left\langle\chi_{MS}(12)|\chi_{MS}(23)\right\rangle=-\frac{1}{2},
\end{aligned}
\label{eq:chiSchiS}
\end{equation}
\begin{equation}
\begin{aligned}
\left\langle\chi_{MS}(12)|\chi_{MA}(23)\right\rangle=-\frac{\sqrt{3}}{2}.
\end{aligned}
\label{eq:chiSchiA}
\end{equation}
Besides, the quark operators of the exchanged neutral light vector mesons for the vertex $VBB$ are given by
\begin{equation}
\begin{aligned}
\tilde{\mathcal{L}}_{V B B} \equiv g\left\{\begin{array}{ll}
\frac{1}{\sqrt{2}}(u \bar{u}-d \bar{d}), & \rho^0 \\
\frac{1}{\sqrt{2}}(u \bar{u}+d \bar{d}), & \omega \\
s \bar{s}, & \phi
\end{array}\right\}.
\end{aligned}
\label{eq:LLVBB}
\end{equation}

Based on the local hidden gauge Lagrangians for the $VPP$ and $VVV$ vertices discussed above, and the extended approach for the $VBB$ vertex with the baryon wave functions, we obtain the general form of the tree level potentials after projecting to the $S$ wave
\begin{equation}
\begin{aligned}
v_{ij}=-C_{ij}\frac{1}{4f_{\pi}^{2}}(p_{i}^{0}+p_{j}^{0}),
\end{aligned}
\label{eq:vij1}
\end{equation}
where $p_{i}^{0}$ and $p_{j}^{0}$ are the energies of the initial and final mesons, respectively, and the coefficient matrices are symmetric, i.e., $C_{ij}=C_{ji}$, which are given in Table \ref{tab:Coefficients}.
In our work, we use a different expression with the relativistic correction \cite{Oset:2001cn}
\begin{equation}
\begin{aligned}
v_{ij}(\sqrt{s})=-C_{ij}\frac{2\sqrt{s}-M_{i}-M_{j}}{4f_{\pi}^{2}}\left(\frac{M_{i}+E_{i}}{2M_{i}}\right)^{1/2}\left(\frac{M_{j}+E_{j}}{2M_{j}}\right)^{1/2},
\end{aligned}
\label{eq:vij2}
\end{equation}
where $M_{i}$ and $M_{j}$ are the mass of the initial and final baryons, respectively, and $E_{i}$, $E_{j}$ their corresponding energies. 
It is worth mentioning that we take $\gamma^{\mu}\approx \gamma^{0}$ in Eq.~\eqref{eq:vij2}, since the transferred momentum is very small when we consider the interaction near the threshold.
More discussions can be found in the Appendix of Ref. \cite{Debastiani:2017ewu}.

\begin{table*}[htbp]
\centering
\renewcommand\tabcolsep{2mm}
\renewcommand{\arraystretch}{1.50}
\caption{The coefficients of the matrix elements $C_{ij}$ with isospin $I=0$ in all systems.}
\begin{tabular*}{178mm}{@{\extracolsep{\fill}}l|cccccccccc}
\toprule[1.00pt]
\toprule[1.00pt]
&\multicolumn{5}{c}{\mbox{$\Omega_{ccc}$-like sector}}&\multicolumn{5}{c}{\mbox{$\Omega_{bbb}$-like sector}}\\
\hline
\multirow{3}{*}{\centering$C_{ij}$$(PB(\frac{1}{2}^{+}))$}&&$D\Xi_{cc}$&$D_{s}\Omega_{cc}$&&&&$\bar{B}\Xi_{bb}$&$\bar{B}_{s}\Omega_{bb}$&&\\
&$D\Xi_{cc}$&$2-\lambda_{cc}$&$\sqrt{2}$&&&$\bar{B}\Xi_{bb}$&$2$&$\sqrt{2}$&&\\
&$D_{s}\Omega_{cc}$&$\sqrt{2}$&$1-\lambda_{cc}$&&&$\bar{B}_{s}\Omega_{bb}$&$\sqrt{2}$&$1$&&\\
\hline
\multirow{5}{*}{\centering$C_{ij}$$(PB(\frac{3}{2}^{+}))$}&&$\eta\Omega_{ccc}$&$\eta^{'}\Omega_{ccc}$&$D\Xi_{cc}^{*}$&$D_{s}\Omega_{cc}^{*}$&&$\eta\Omega_{bbb}$&$\eta^{'}\Omega_{bbb}$&$\bar{B}\Xi_{bb}^{*}$&$\bar{B}_{s}\Omega_{bb}^{*}$\\
&$\eta\Omega_{ccc}$&$0$&$0$&$\frac{-\sqrt{2}\lambda_{c}}{\sqrt{3}}$&$\frac{\lambda_{c}}{\sqrt{3}}$&$\eta\Omega_{bbb}$&$0$&$0$&$\frac{-\sqrt{2}\lambda_{b}}{\sqrt{3}}$&$\frac{\lambda_{b}}{\sqrt{3}}$\\
&$\eta^{'}\Omega_{ccc}$&$0$&$0$&$\frac{-\lambda_{c}}{\sqrt{3}}$&$\frac{-\sqrt{2}\lambda_{c}}{\sqrt{3}}$&$\eta^{'}\Omega_{bbb}$&$0$&$0$&$\frac{-\lambda_{b}}{\sqrt{3}}$&$\frac{-\sqrt{2}\lambda_{b}}{\sqrt{3}}$\\
&$D\Xi_{cc}^{*}$&$\frac{-\sqrt{2}\lambda_{c}}{\sqrt{3}}$&$\frac{-\lambda_{c}}{\sqrt{3}}$&$2-\lambda_{cc}$&$\sqrt{2}$&$\bar{B}\Xi_{bb}^{*}$&$\frac{-\sqrt{2}\lambda_{b}}{\sqrt{3}}$&$\frac{-\lambda_{b}}{\sqrt{3}}$&$2$&$\sqrt{2}$\\
&$D_{s}\Omega_{cc}^{*}$&$\frac{\lambda_{c}}{\sqrt{3}}$&$\frac{-\sqrt{2}\lambda_{c}}{\sqrt{3}}$&$\sqrt{2}$&$1-\lambda_{cc}$&$\bar{B}_{s}\Omega_{bb}^{*}$&$\frac{\lambda_{b}}{\sqrt{3}}$&$\frac{-\sqrt{2}\lambda_{b}}{\sqrt{3}}$&$\sqrt{2}$&$1$\\
\hline
\multirow{3}{*}{\centering$C_{ij}$$(VB(\frac{1}{2}^{+}))$}&&$D^{*}\Xi_{cc}$&$D_{s}^{*}\Omega_{cc}$&&&&$\bar{B}^{*}\Xi_{bb}$&$\bar{B}_{s}^{*}\Omega_{bb}$&&\\
&$D^{*}\Xi_{cc}$&$2-\lambda_{cc}$&$\sqrt{2}$&&&$\bar{B}^{*}\Xi_{bb}$&$2$&$\sqrt{2}$&&\\
&$D_{s}^{*}\Omega_{cc}$&$\sqrt{2}$&$1-\lambda_{cc}$&&&$\bar{B}_{s}^{*}\Omega_{bb}$&$\sqrt{2}$&$1$&&\\
\hline
\multirow{5}{*}{\centering$C_{ij}$$(VB(\frac{3}{2}^{+}))$}&&$\omega\Omega_{ccc}$&$\phi\Omega_{ccc}$&$D^{*}\Xi_{cc}^{*}$&$D_{s}^{*}\Omega_{cc}^{*}$&&$\omega\Omega_{bbb}$&$\phi\Omega_{bbb}$&$\bar{B}^{*}\Xi_{bb}^{*}$&$\bar{B}_{s}^{*}\Omega_{bb}^{*}$\\
&$\omega\Omega_{ccc}$&$0$&$0$&$-\lambda_{c}$&$0$&$\omega\Omega_{bbb}$&$0$&$0$&$-\lambda_{b}$&$0$\\
&$\phi\Omega_{ccc}$&$0$&$0$&$0$&$-\lambda_{c}$&$\phi\Omega_{bbb}$&$0$&$0$&$0$&$-\lambda_{b}$\\
&$D^{*}\Xi_{cc}^{*}$&$-\lambda_{c}$&$0$&$2-\lambda_{cc}$&$\sqrt{2}$&$\bar{B}^{*}\Xi_{bb}^{*}$&$-\lambda_{b}$&$0$&$2$&$\sqrt{2}$\\
&$D_{s}^{*}\Omega_{cc}^{*}$&$0$&$-\lambda_{c}$&$\sqrt{2}$&$1-\lambda_{cc}$&$\bar{B}_{s}^{*}\Omega_{bb}^{*}$&$0$&$-\lambda_{b}$&$\sqrt{2}$&$1$\\
\hline
&\multicolumn{10}{c}{\mbox{$\Omega_{bcc}$-like sector}}\\
\hline
\multirow{9}{*}{\centering$C_{ij}$$(PB(\frac{1}{2}^{+}))$}&&$\eta\Omega_{bcc}$&$\eta^{'}\Omega_{bcc}$&$D\Xi_{bc}$&$D\Xi_{bc}^{'}$&$\bar{B}\Xi_{cc}$&$D_{s}\Omega_{bc}$&$D_{s}\Omega_{bc}^{'}$&$\bar{B}_{s}\Omega_{cc}$&\\
&$\eta\Omega_{bcc}$&$0$&$0$&$0$&$\frac{-2\lambda_{c}}{\sqrt{3}}$&$\frac{-\sqrt{2}\lambda_{b}}{\sqrt{3}}$&$0$&$\frac{\sqrt{2}\lambda_{c}}{\sqrt{3}}$&$\frac{\lambda_{b}}{\sqrt{3}}$&\\
&$\eta^{'}\Omega_{bcc}$&$0$&$0$&$0$&$\frac{-\sqrt{2}\lambda_{c}}{\sqrt{3}}$&$\frac{-\lambda_{b}}{\sqrt{3}}$&$0$&$\frac{-2\lambda_{c}}{\sqrt{3}}$&$\frac{-\sqrt{2}\lambda_{b}}{\sqrt{3}}$&\\
&$D\Xi_{bc}$&$0$&$0$&$2-\lambda_{cc}$&$0$&$0$&$\sqrt{2}$&$0$&$0$&\\
&$D\Xi_{bc}^{'}$&$\frac{-2\lambda_{c}}{\sqrt{3}}$&$\frac{-\sqrt{2}\lambda_{c}}{\sqrt{3}}$&$0$&$2-\lambda_{cc}$&$0$&$0$&$\sqrt{2}$&$0$&\\
&$\bar{B}\Xi_{cc}$&$\frac{-\sqrt{2}\lambda_{b}}{\sqrt{3}}$&$\frac{-\lambda_{b}}{\sqrt{3}}$&$0$&$0$&$2$&$0$&$0$&$\sqrt{2}$&\\
&$D_{s}\Omega_{bc}$&$0$&$0$&$\sqrt{2}$&$0$&$0$&$1-\lambda_{cc}$&$0$&$0$&\\
&$D_{s}\Omega_{bc}^{'}$&$\frac{\sqrt{2}\lambda_{c}}{\sqrt{3}}$&$\frac{-2\lambda_{c}}{\sqrt{3}}$&$0$&$\sqrt{2}$&$0$&$0$&$1-\lambda_{cc}$&$0$&\\
&$\bar{B}_{s}\Omega_{cc}$&$\frac{\lambda_{b}}{\sqrt{3}}$&$\frac{-\sqrt{2}\lambda_{b}}{\sqrt{3}}$&$0$&$0$&$\sqrt{2}$&$0$&$0$&$1$&\\
\hline
\multirow{7}{*}{\centering$C_{ij}$$(PB(\frac{3}{2}^{+}))$}&&$\eta\Omega_{bcc}^{*}$&$\eta^{'}\Omega_{bcc}^{*}$&$D\Xi_{bc}^{*}$&$\bar{B}\Xi_{cc}^{*}$&$D_{s}\Omega_{bc}^{*}$&$\bar{B}_{s}\Omega_{cc}^{*}$&&&\\
&$\eta\Omega_{bcc}^{*}$&$0$&$0$&$\frac{-2\lambda_{c}}{\sqrt{3}}$&$\frac{-\sqrt{2}\lambda_{b}}{\sqrt{3}}$&$\frac{\sqrt{2}\lambda_{c}}{\sqrt{3}}$&$\frac{\lambda_{b}}{\sqrt{3}}$&&&\\
&$\eta^{'}\Omega_{bcc}^{*}$&$0$&$0$&$\frac{-\sqrt{2}\lambda_{c}}{\sqrt{3}}$&$\frac{-\lambda_{b}}{\sqrt{3}}$&$\frac{-2\lambda_{c}}{\sqrt{3}}$&$\frac{-\sqrt{2}\lambda_{b}}{\sqrt{3}}$&&&\\
&$D\Xi_{bc}^{*}$&$\frac{-2\lambda_{c}}{\sqrt{3}}$&$\frac{-\sqrt{2}\lambda_{c}}{\sqrt{3}}$&$2-\lambda_{cc}$&$0$&$\sqrt{2}$&$0$&&&\\
&$\bar{B}\Xi_{cc}^{*}$&$\frac{-\sqrt{2}\lambda_{b}}{\sqrt{3}}$&$\frac{-\lambda_{b}}{\sqrt{3}}$&$0$&$2$&$0$&$\sqrt{2}$&&&\\
&$D_{s}\Omega_{bc}^{*}$&$\frac{\sqrt{2}\lambda_{c}}{\sqrt{3}}$&$\frac{-2\lambda_{c}}{\sqrt{3}}$&$\sqrt{2}$&$0$&$1-\lambda_{cc}$&$0$&&&\\
&$\bar{B}_{s}\Omega_{cc}^{*}$&$\frac{\lambda_{b}}{\sqrt{3}}$&$\frac{-\sqrt{2}\lambda_{b}}{\sqrt{3}}$&$0$&$\sqrt{2}$&$0$&$1$&&&\\
\hline
\multirow{9}{*}{\centering$C_{ij}$$(VB(\frac{1}{2}^{+}))$}&&$\omega\Omega_{bcc}$&$\phi\Omega_{bcc}$&$D^{*}\Xi_{bc}$&$D^{*}\Xi_{bc}^{'}$&$\bar{B}^{*}\Xi_{cc}$&$D_{s}^{*}\Omega_{bc}$&$D_{s}^{*}\Omega_{bc}^{'}$&$\bar{B}_{s}^{*}\Omega_{cc}$&\\
&$\omega\Omega_{bcc}$&$0$&$0$&$0$&$-\sqrt{2}\lambda_{c}$&$-\lambda_{b}$&$0$&$0$&$0$&\\
&$\phi\Omega_{bcc}$&$0$&$0$&$0$&$0$&$0$&$0$&$-\sqrt{2}\lambda_{c}$&$-\lambda_{b}$&\\
&$D^{*}\Xi_{bc}$&$0$&$0$&$2-\lambda_{cc}$&$0$&$0$&$\sqrt{2}$&$0$&$0$&\\
&$D^{*}\Xi_{bc}^{'}$&$-\sqrt{2}\lambda_{c}$&$0$&$0$&$2-\lambda_{cc}$&$0$&$0$&$\sqrt{2}$&$0$&\\
&$\bar{B}^{*}\Xi_{cc}$&$-\lambda_{b}$&$0$&$0$&$0$&$2$&$0$&$0$&$\sqrt{2}$&\\
&$D_{s}^{*}\Omega_{bc}$&$0$&$0$&$\sqrt{2}$&$0$&$0$&$1-\lambda_{cc}$&$0$&$0$&\\
&$D_{s}^{*}\Omega_{bc}^{'}$&$0$&$-\sqrt{2}\lambda_{c}$&$0$&$\sqrt{2}$&$0$&$0$&$1-\lambda_{cc}$&$0$&\\
&$\bar{B}_{s}^{*}\Omega_{cc}$&$0$&$-\lambda_{b}$&$0$&$0$&$\sqrt{2}$&$0$&$0$&$1$&\\
\hline
\multicolumn{1}{c}{{\it $-$Continue.$-$}}
\end{tabular*}
\label{tab:Coefficients}
\end{table*}

\begin{table*}[htbp]
\centering
\renewcommand\tabcolsep{2mm}
\renewcommand{\arraystretch}{1.50}
\begin{tabular*}{178mm}{@{\extracolsep{\fill}}l|cccccccccc}
\multicolumn{1}{c}{{\it $-$Continue.$-$}}\\
\hline
\multirow{7}{*}{\centering$C_{ij}$$(VB(\frac{3}{2}^{+}))$}&&$\omega\Omega_{bcc}^{*}$&$\phi\Omega_{bcc}^{*}$&$D^{*}\Xi_{bc}^{*}$&$\bar{B}^{*}\Xi_{cc}^{*}$&$D_{s}^{*}\Omega_{bc}^{*}$&$\bar{B}_{s}^{*}\Omega_{cc}^{*}$&&&\\
&$\omega\Omega_{bcc}^{*}$&$0$&$0$&$-\sqrt{2}\lambda_{c}$&$-\lambda_{b}$&$0$&$0$&&&\\
&$\phi\Omega_{bcc}^{*}$&$0$&$0$&$0$&$0$&$-\sqrt{2}\lambda_{c}$&$-\lambda_{b}$&&&\\
&$D^{*}\Xi_{bc}^{*}$&$-\sqrt{2}\lambda_{c}$&$0$&$2-\lambda_{cc}$&$0$&$\sqrt{2}$&$0$&&&\\
&$\bar{B}^{*}\Xi_{cc}^{*}$&$-\lambda_{b}$&$0$&$0$&$2$&$0$&$\sqrt{2}$&&&\\
&$D_{s}^{*}\Omega_{bc}^{*}$&$0$&$-\sqrt{2}\lambda_{c}$&$\sqrt{2}$&$0$&$1-\lambda_{cc}$&$0$&&&\\
&$\bar{B}_{s}^{*}\Omega_{cc}^{*}$&$0$&$-\lambda_{b}$&$0$&$\sqrt{2}$&$0$&$1$&&&\\
\hline
&\multicolumn{10}{c}{\mbox{$\Omega_{bbc}$-like sector}}\\
\hline
\multirow{9}{*}{\centering$C_{ij}$$(PB(\frac{1}{2}^{+}))$}&&$\eta\Omega_{bbc}$&$\eta^{'}\Omega_{bbc}$&$D\Xi_{bb}$&$\bar{B}\Xi_{bc}$&$\bar{B}\Xi_{bc}^{'}$&$D_{s}\Omega_{bb}$&$\bar{B}_{s}\Omega_{bc}$&$\bar{B}_{s}\Omega_{bc}^{'}$&\\
&$\eta\Omega_{bbc}$&$0$&$0$&$\frac{-\sqrt{2}\lambda_{c}}{\sqrt{3}}$&$\frac{-\lambda_{b}}{2}$&$\frac{\lambda_{b}}{2\sqrt{3}}$&$\frac{\lambda_{c}}{\sqrt{3}}$&$\frac{\lambda_{b}}{2\sqrt{2}}$&$\frac{-\lambda_{b}}{2\sqrt{6}}$&\\
&$\eta^{'}\Omega_{bbc}$&$0$&$0$&$\frac{-\lambda_{c}}{\sqrt{3}}$&$\frac{-\lambda_{b}}{2\sqrt{2}}$&$\frac{\lambda_{b}}{2\sqrt{6}}$&$\frac{-\sqrt{2}\lambda_{c}}{\sqrt{3}}$&$\frac{-\lambda_{b}}{2}$&$\frac{\lambda_{b}}{2\sqrt{3}}$&\\
&$D\Xi_{bb}$&$\frac{-\sqrt{2}\lambda_{c}}{\sqrt{3}}$&$\frac{-\lambda_{c}}{\sqrt{3}}$&$2$&$0$&$0$&$\sqrt{2}$&$0$&$0$&\\
&$\bar{B}\Xi_{bc}$&$\frac{-\lambda_{b}}{2}$&$\frac{-\lambda_{b}}{2\sqrt{2}}$&$0$&$2$&$0$&$0$&$\sqrt{2}$&$0$&\\
&$\bar{B}\Xi_{bc}^{'}$&$\frac{\lambda_{b}}{2\sqrt{3}}$&$\frac{\lambda_{b}}{2\sqrt{6}}$&$0$&$0$&$2$&$0$&$0$&$\sqrt{2}$&\\
&$D_{s}\Omega_{bb}$&$\frac{\lambda_{c}}{\sqrt{3}}$&$\frac{-\sqrt{2}\lambda_{c}}{\sqrt{3}}$&$\sqrt{2}$&$0$&$0$&$1$&$0$&$0$&\\
&$\bar{B}_{s}\Omega_{bc}$&$\frac{\lambda_{b}}{2\sqrt{2}}$&$\frac{-\lambda_{b}}{2}$&$0$&$\sqrt{2}$&$0$&$0$&$1$&$0$&\\
&$\bar{B}_{s}\Omega_{bc}^{'}$&$\frac{-\lambda_{b}}{2\sqrt{6}}$&$\frac{\lambda_{b}}{2\sqrt{3}}$&$0$&$0$&$\sqrt{2}$&$0$&$0$&$1$&\\
\hline
\multirow{7}{*}{\centering$C_{ij}$$(PB(\frac{3}{2}^{+}))$}&&$\eta\Omega_{bbc}^{*}$&$\eta^{'}\Omega_{bbc}^{*}$&$D\Xi_{bb}^{*}$&$\bar{B}\Xi_{bc}^{*}$&$D_{s}\Omega_{bb}^{*}$&$\bar{B}_{s}\Omega_{bc}^{*}$&&&\\
&$\eta\Omega_{bbc}^{*}$&$0$&$0$&$\frac{-\sqrt{2}\lambda_{c}}{\sqrt{3}}$&$\frac{-\lambda_{b}}{\sqrt{3}}$&$\frac{\lambda_{c}}{\sqrt{3}}$&$\frac{\lambda_{b}}{\sqrt{6}}$&&&\\
&$\eta^{'}\Omega_{bbc}^{*}$&$0$&$0$&$\frac{-\lambda_{c}}{\sqrt{3}}$&$\frac{-\lambda_{b}}{\sqrt{6}}$&$\frac{-\sqrt{2}\lambda_{c}}{\sqrt{3}}$&$\frac{-\lambda_{b}}{\sqrt{3}}$&&&\\
&$D\Xi_{bb}^{*}$&$\frac{-\sqrt{2}\lambda_{c}}{\sqrt{3}}$&$\frac{-\lambda_{c}}{\sqrt{3}}$&$2$&$0$&$\sqrt{2}$&$0$&&&\\
&$\bar{B}\Xi_{bc}^{*}$&$\frac{-\lambda_{b}}{\sqrt{3}}$&$\frac{-\lambda_{b}}{\sqrt{6}}$&$0$&$2$&$0$&$\sqrt{2}$&&&\\
&$D_{s}\Omega_{bb}^{*}$&$\frac{\lambda_{c}}{\sqrt{3}}$&$\frac{-\sqrt{2}\lambda_{c}}{\sqrt{3}}$&$\sqrt{2}$&$0$&$1$&$0$&&&\\
&$\bar{B}_{s}\Omega_{bc}^{*}$&$\frac{\lambda_{b}}{\sqrt{6}}$&$\frac{-\lambda_{b}}{\sqrt{3}}$&$0$&$\sqrt{2}$&$0$&$1$&&&\\
\hline
\multirow{9}{*}{\centering$C_{ij}$$(VB(\frac{1}{2}^{+}))$}&&$\omega\Omega_{bbc}$&$\phi\Omega_{bbc}$&$D^{*}\Xi_{bb}$&$\bar{B}^{*}\Xi_{bc}$&$\bar{B}^{*}\Xi_{bc}^{'}$&$D_{s}^{*}\Omega_{bb}$&$\bar{B}_{s}^{*}\Omega_{bc}$&$\bar{B}_{s}^{*}\Omega_{bc}^{'}$&\\
&$\omega\Omega_{bbc}$&$0$&$0$&$-\lambda_{c}$&$\frac{-\sqrt{3}\lambda_{b}}{2\sqrt{2}}$&$\frac{\lambda_{b}}{2\sqrt{2}}$&$0$&$0$&$0$&\\
&$\phi\Omega_{bbc}$&$0$&$0$&$0$&$0$&$0$&$-\lambda_{c}$&$\frac{-\sqrt{3}\lambda_{b}}{2\sqrt{2}}$&$\frac{\lambda_{b}}{2\sqrt{2}}$&\\
&$D^{*}\Xi_{bb}$&$-\lambda_{c}$&$0$&$2$&$0$&$0$&$\sqrt{2}$&$0$&$0$&\\
&$\bar{B}^{*}\Xi_{bc}$&$\frac{-\sqrt{3}\lambda_{b}}{2\sqrt{2}}$&$0$&$0$&$2$&$0$&$0$&$\sqrt{2}$&$0$&\\
&$\bar{B}^{*}\Xi_{bc}^{'}$&$\frac{\lambda_{b}}{2\sqrt{2}}$&$0$&$0$&$0$&$2$&$0$&$0$&$\sqrt{2}$&\\
&$D_{s}^{*}\Omega_{bb}$&$0$&$-\lambda_{c}$&$\sqrt{2}$&$0$&$0$&$1$&$0$&$0$&\\
&$\bar{B}_{s}^{*}\Omega_{bc}$&$0$&$\frac{-\sqrt{3}\lambda_{b}}{2\sqrt{2}}$&$0$&$\sqrt{2}$&$0$&$0$&$1$&$0$&\\
&$\bar{B}_{s}^{*}\Omega_{bc}^{'}$&$0$&$\frac{\lambda_{b}}{2\sqrt{2}}$&$0$&$0$&$\sqrt{2}$&$0$&$0$&$1$&\\
\hline
\multirow{7}{*}{\centering$C_{ij}$$(VB(\frac{3}{2}^{+}))$}&&$\omega\Omega_{bbc}^{*}$&$\phi\Omega_{bbc}^{*}$&$D^{*}\Xi_{bb}^{*}$&$\bar{B}^{*}\Xi_{bc}^{*}$&$D_{s}^{*}\Omega_{bb}^{*}$&$\bar{B}_{s}^{*}\Omega_{bc}^{*}$&&&\\
&$\omega\Omega_{bbc}^{*}$&$0$&$0$&$-\lambda_{c}$&$\frac{-\lambda_{b}}{\sqrt{2}}$&$0$&$0$&&&\\
&$\phi\Omega_{bbc}^{*}$&$0$&$0$&$0$&$0$&$-\lambda_{c}$&$\frac{-\lambda_{b}}{\sqrt{2}}$&&&\\
&$D^{*}\Xi_{bb}^{*}$&$-\lambda_{c}$&$0$&$2$&$0$&$\sqrt{2}$&$0$&&&\\
&$\bar{B}^{*}\Xi_{bc}^{*}$&$\frac{-\lambda_{b}}{\sqrt{2}}$&$0$&$0$&$2$&$0$&$\sqrt{2}$&&&\\
&$D_{s}^{*}\Omega_{bb}^{*}$&$0$&$-\lambda_{c}$&$\sqrt{2}$&$0$&$1$&$0$&&&\\
&$\bar{B}_{s}^{*}\Omega_{bc}^{*}$&$0$&$\frac{-\lambda_{b}}{\sqrt{2}}$&$0$&$\sqrt{2}$&$0$&$1$&&&\\
\bottomrule[1.00pt]
\bottomrule[1.00pt]
\end{tabular*}
\label{tab:Coefficients_Continued}
\end{table*}

Note that the coefficients of $C_{ij}$ in Table \ref{tab:Coefficients} are consistent with our isospin notation, where we use the phase convention $|D^{(*)+}\rangle=| 1/2,1/2\rangle$, $|D^{(*)0}\rangle=-| 1/2,-1/2\rangle$, $|\bar{B}^{(*)0}\rangle=| 1/2,1/2\rangle$, and $|B^{(*)-}\rangle=-| 1/2,-1/2\rangle$.
Here, we only investigate the isospin $I=0$ sectors, since the interaction potentials of the $I=1$ sectors are repulsive for these systems and bound states do not appear. 
In addition, the three parameters in Table \ref{tab:Coefficients} are taken as $\lambda_{c}\approx 1/4$ \cite{Debastiani:2017ewu,Wang:2022aga}, $\lambda_{cc}\approx 1/9$ \cite{Marse-Valera:2022khy}, and $\lambda_{b}\approx 1/10$ \cite{Dias:2019klk}, which are the suppression factors of the exchanged heavy vector mesons, $D_{(s)}^{*}$, $J/\psi$, and $B_{(s)}^{*}$ with respect to the exchanged light vector mesons, respectively.
The contributions from the exchange of heavier vector mesons, such as $B_{c}^{*}$ and $\Upsilon$, are neglected.
It is significant to note that the dominant contributions come from exchanging the light vector mesons.
More discussions can be found in Refs. \cite{Debastiani:2017ewu,Roca:2024nsi}.

With the interaction potentials of the $S$ wave obtained above, the scattering amplitude can be calculated by solving the coupled channel Bethe-Salpeter equation with the on-shell description~\cite{Oset:1997it}
\begin{equation}
\begin{aligned} 
T = [1-vG]^{-1}v,
\end{aligned}
\label{eq:BSE}
\end{equation}
where $G$ is the diagonal matrix consisting of the elements for the meson-baryon loop functions
\begin{equation}
\begin{aligned} 
G_{l}=i \int \frac{d^{4} q}{(2 \pi)^{4}}\frac{2M_{l}}{(P-q)^{2}-M_{l}^{2}+i \epsilon} \frac{1}{q^{2}-m_{l}^{2}+i \epsilon}.
\end{aligned}
\label{eq:G}
\end{equation}
This magnitude is divergent and we can use the three-momentum cutoff approach \cite{Oset:1997it} or the dimensional regularization method \cite{Oller:2000fj,Jido:2003cb}.
The expression for the three-momentum cutoff approach is given by
\begin{equation}
\begin{aligned}
G_{l}(s)=\int_{0}^{q_{max}} \frac{\vec{q}\,^{2} d \vec{q}}{2 \pi^{2}} \frac{1}{2 \omega_{l}(\vec{q})} \frac{M_{l}}{E_{l}(\vec{q})} \frac{1}{p^{0}+k^{0}-\omega_{l}(\vec{q})-E_{l}(\vec{q})+i \epsilon},
\end{aligned}
\label{eq:GCO}
\end{equation}
where $p^{0}+k^{0}=\sqrt{s}$, $\omega_{l}(\vec{q})=\sqrt{\vec{q}\,^2+m_{l}^2}$, and $E_{l}(\vec{q})=\sqrt{\vec{q}\,^2+M_{l}^2}$, with $m_{l}$ and $M_{l}$ the masses of meson and baryon of the $l$ channel, respectively, and $q_{max}$ is the only free parameter. 
Furthermore, the formula for the dimensional regularization method is given by
\begin{equation}
\begin{aligned}
G_{l}(s)=& \frac{2 M_{l}}{16 \pi^{2}}\left\{a(\mu)+\ln \frac{M_{l}^{2}}{\mu^{2}}+\frac{m_{l}^{2}-M_{l}^{2}+s}{2 s} \ln \frac{m_{l}^{2}}{M_{l}^{2}}\right.\\
&+\frac{q_{cml}(s)}{\sqrt{s}}\left[\ln \left(s-\left(M_{l}^{2}-m_{l}^{2}\right)+2 q_{cml}(s) \sqrt{s}\right)\right.\\
&+\ln \left(s+\left(M_{l}^{2}-m_{l}^{2}\right)+2 q_{cml}(s) \sqrt{s}\right) \\
&-\ln \left(-s-\left(M_{l}^{2}-m_{l}^{2}\right)+2 q_{cml}(s) \sqrt{s}\right) \\
&\left.\left.-\ln \left(-s+\left(M_{l}^{2}-m_{l}^{2}\right)+2 q_{cml}(s) \sqrt{s}\right)\right]\right\},
\end{aligned}
\label{eq:GDR}
\end{equation}
which also has only one free parameter the regularization scale $\mu$, and the subtraction constant $a(\mu)$ depends on the $\mu$ chosen, and where $q_{cml}(s)$ is the three momentum of the particle in the center-of-mass frame
\begin{equation}
\begin{aligned}
q_{cml}(s)=\frac{\lambda^{1 / 2}\left(s, M_{l}^{2}, m_{l}^{2}\right)}{2 \sqrt{s}},
\end{aligned}
\end{equation}
with the K\"all\'en triangle function $\lambda(a, b, c)=a^{2}+b^{2}+c^{2}-2(a b+a c+b c)$.
In the present work, we take the dimensional regularization method to regularize Eq. (\ref{eq:G}).
The values of the regularization scale $\mu$ and the subtraction constant $a(\mu)$ are discussed in detail in the next section.

Within the coupled channel framework, we first search for peak structures in the $T_{ij}$ scattering amplitudes, and then determine the masses and widths of the resonances by searching for the poles of the scattering amplitudes on the complex Riemann sheets.
If a channel is open for the decay channel, it is necessary to extrapolate the loop function $G_{l}(s)$ to the second Riemann sheet by the continuous condition
\begin{equation}
\begin{aligned}
G_{l}^{(II)}(s)&=G_{l}(s)-2i \text{Im}G_{l}(s)\\
&=G_{l}(s)+\frac{i}{2\pi}\frac{M_{l}q_{cml}(s)}{\sqrt{s}}.
\end{aligned}
\end{equation}
Furthermore, we can evaluate the couplings of the generated state for a given channel by performing a Laurent expansion on the amplitude in the vicinity of the pole $\sqrt{s_{p}}$ \cite{Yamagata-Sekihara:2010kpd}
\begin{equation}
T_{ij}=\frac{g_{i}g_{j}}{\sqrt{s}-\sqrt{s_{p}}}.
\end{equation}
In addition, the Weinberg’s rule \cite{Weinberg:1965zz} for the bound state or resonance can be generalized to the formalism of the coupled channels approach with the couplings \cite{Aceti:2012dd}
\begin{equation}
-\sum_{i} g_{i}^{2}\left[\frac{d G_{l}}{d \sqrt{s}}\right]_{\sqrt{s}=\sqrt{s_{p}}}=1,
\label{eq:sumrule}
\end{equation}
which is applied for the purely molecular states, where each of the terms on the left hand side of Eq. \eqref{eq:sumrule} gives the probability of each channel. In the case of composite states, where the states contain not only the molecular components but also the other non-molecular components, this sum rule of Eq.~\eqref{eq:sumrule} can be modified as~\cite{Aceti:2012dd}
\begin{equation}
-\sum_{i} g_{i}^{2}\left[\frac{d G_{l}}{d \sqrt{s}}\right]_{\sqrt{s}=\sqrt{s_{p}}}=1-Z,
\label{eq:components}
\end{equation}
where $Z$ stands for the other non-molecular components in the bound state or resonance.

\section{Results}\label{sec:Results}

As discussed in the last section, there are only one free parameter in our formalism, the regularization scale $\mu$, and the subtraction constant $a(\mu)$ depends on the $\mu$ chosen. 
Thus, we should first determine the values of the regularization scale $\mu$ and the subtraction constant $a(\mu)$ in the loop functions. 
At first, we take $\mu=q_{max}=800$ MeV~\cite{Marse-Valera:2022khy,Roca:2024nsi}, and then use the three-momentum cutoff approach and the dimensional regularization method to match their values of the loop function at the threshold for a given channel to determine $a(\mu)$, as done in Ref. \cite{Oset:2001cn}.

In the $\Omega_{ccc}$-like sector, we find four candidates for the three-charm molecular pentaquarks, as shown in Table \ref{tab:Poles_ccc}, where the corresponding poles, couplings and compositeness for a given coupled channel are presented.
For each pole, we use the $+$ to indicate that the corresponding coupled channel is closed, and the $-$ to specify the open channel accordingly, which is the decay channel.
The largest coupling and compositeness are shown in bold, which indicates the most relevant channel in most cases, in other words, the most relevant bound channel.
Thus, the first two poles in Table \ref{tab:Poles_ccc}, $5446.97$ MeV and $(5493.65 - 5.92i)$ MeV, are mainly coupled to the channels $D\Xi_{cc}$ and $D\Xi_{cc}^{*}$, respectively, indicating that they are the molecular states of these two channels with the binding energies of $42$ MeV and $49$ MeV, respectively, while the contributions of the channels $D_{s}\Omega_{cc}$ and $D_{s}\Omega_{cc}^{*}$ are also significant with large couplings and compositenesses.
And the second state has a width of about $\Gamma=12$ MeV, coming from the decay into the $\eta\Omega_{ccc}$ channel. 
The last two poles, $5591.87$ MeV and $(5640.10 - 3.43i)$ MeV, are degenerate in $J^{P}=\frac{1}{2}^{-}$, $\frac{3}{2}^{-}$ and $J^{P}=\frac{1}{2}^{-}$, $\frac{3}{2}^{-}$, $\frac{5}{2}^{-}$, respectively, which are possibly qualified as the bound states of $D^{*}\Xi_{cc}$ and $D^{*}\Xi_{cc}^{*}$ with binding energies of $39$ MeV and $43$ MeV, respectively.
One can see that the last state can decay into the $\omega\Omega_{ccc}$ channel with a width of about $\Gamma=7$ MeV.
From the perspective of compositeness results, our predicted states are mostly pure molecular states with very small non-molecular components, since the sum of the compositenesses is nearly close to one.

\begin{table}[htbp]
\centering
\renewcommand\tabcolsep{2.0mm}
\renewcommand{\arraystretch}{1.50}
\caption{The poles (in MeV), couplings $|g_{i}|$ and compositeness $|1-Z|_{i}$ for each channel in the $\Omega_{ccc}$-like sector.}
\begin{tabular*}{86mm}{@{\extracolsep{\fill}}l|ccccc}
\toprule[1.00pt]
\toprule[1.00pt]
\multicolumn{6}{c}{\mbox{$\Omega_{ccc}$-like sector}}\\
\hline
&Channels&$D\Xi_{cc}$&$D_{s}\Omega_{cc}$&&\\
\hline
\multirow{3}{*}{\centering$0(\frac{1}{2}^{-})$}&\multicolumn{5}{l}{\mbox{$5446.97$ $(++)$}}\\
&$|g_{i}|$&$\textbf{2.13}$&$1.81$&&\\
&$|1-Z|_{i}$&$\textbf{0.79}$&$0.20$&&\\
\hline
&Channels&$\eta\Omega_{ccc}$&$\eta^{'}\Omega_{ccc}$&$D\Xi_{cc}^{*}$&$D_{s}\Omega_{cc}^{*}$\\
\hline
\multirow{3}{*}{\centering$0(\frac{3}{2}^{-})$}&\multicolumn{5}{l}{\mbox{$5493.65-5.92i$ $(-+++)$}}\\
&$|g_{i}|$&$0.46$&$0.01$&$\textbf{2.27}$&$1.59$\\
&$|1-Z|_{i}$&$0.02$&$0.00$&$\textbf{0.83}$&$0.15$\\
\hline
&Channels&$D^{*}\Xi_{cc}$&$D_{s}^{*}\Omega_{cc}$&&\\
\hline
\multirow{3}{*}{\centering$0(\frac{1}{2}^{-},\frac{3}{2}^{-})$}&\multicolumn{5}{l}{\mbox{$5591.87$ $(++)$}}\\
&$|g_{i}|$&$\textbf{2.08}$&$1.81$&&\\
&$|1-Z|_{i}$&$\textbf{0.79}$&$0.20$&&\\
\hline
&Channels&$\omega\Omega_{ccc}$&$\phi\Omega_{ccc}$&$D^{*}\Xi_{cc}^{*}$&$D_{s}^{*}\Omega_{cc}^{*}$\\
\hline
\multirow{3}{*}{\centering$0(\frac{1}{2}^{-},\frac{3}{2}^{-},\frac{5}{2}^{-})$}&\multicolumn{5}{l}{\mbox{$5640.10-3.43i$ $(-+++)$}}\\
&$|g_{i}|$&$0.42$&$0.32$&$\textbf{2.17}$&$1.68$\\
&$|1-Z|_{i}$&$0.03$&$0.01$&$\textbf{0.81}$&$0.17$\\
\bottomrule[1.00pt]
\bottomrule[1.00pt]
\end{tabular*}
\label{tab:Poles_ccc}
\end{table}

\begin{figure*}[htbp]
\begin{minipage}{0.33\linewidth}
\centering
\includegraphics[width=1\linewidth,trim=0 0 0 0,clip]{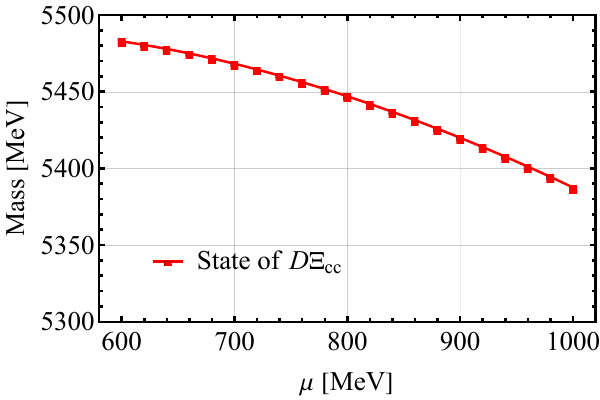} 
\label{fig:ccc_1}
\end{minipage}
\begin{minipage}{0.33\linewidth} 
\centering 
\includegraphics[width=1\linewidth,trim=0 0 0 0,clip]{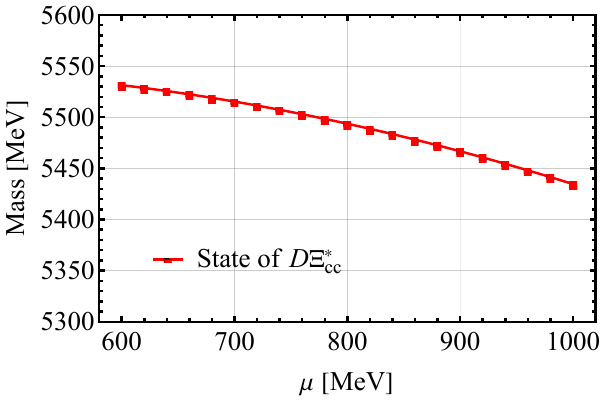} 
\label{fig:ccc_2}  
\end{minipage}	
\begin{minipage}{0.33\linewidth} 
\centering 
\includegraphics[width=1\linewidth,trim=0 0 0 0,clip]{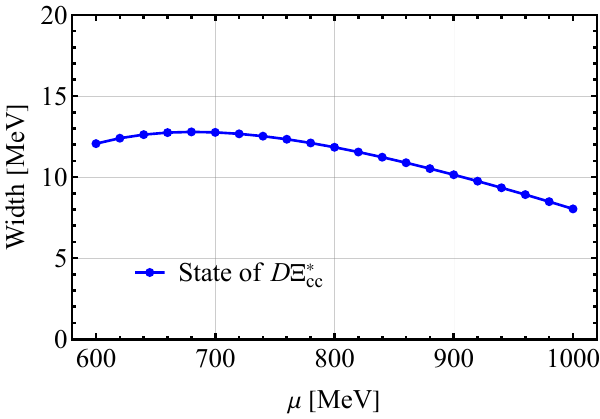} 
\label{fig:ccc_3}  
\end{minipage}
\begin{minipage}{0.33\linewidth} 
\centering 
\includegraphics[width=1\linewidth,trim=0 0 0 0,clip]{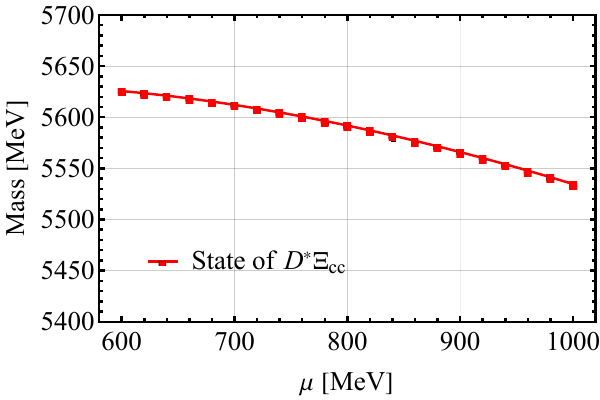} 
\label{fig:ccc_4}  
\end{minipage}
\begin{minipage}{0.33\linewidth} 
\centering 
\includegraphics[width=1\linewidth,trim=0 0 0 0,clip]{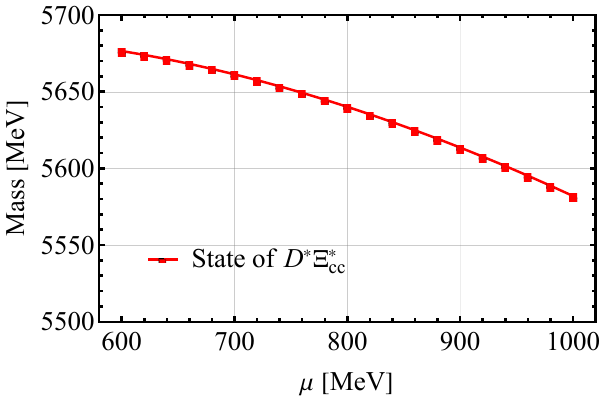} 
\label{fig:ccc_5}  
\end{minipage}
\begin{minipage}{0.33\linewidth} 
\centering 
\includegraphics[width=1\linewidth,trim=0 0 0 0,clip]{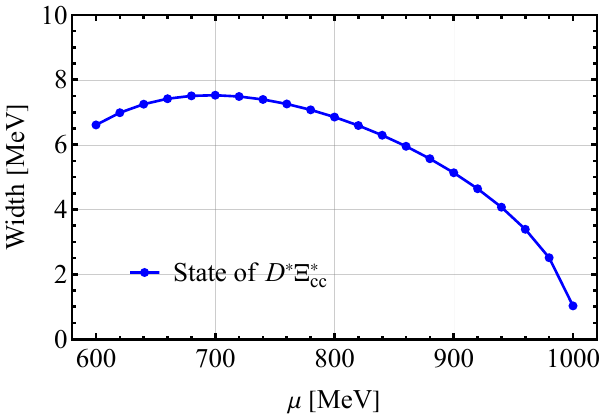} 
\label{fig:ccc_6}  
\end{minipage}
\caption{Trajectories for the masses and the widths of the $\Omega_{ccc}$-like states by varying the regularization scale $\mu$.}
\label{fig:ccc}
\end{figure*}

Furthermore, in order to understand the properties of the bound states, in Fig. \ref{fig:ccc} we also plot the changes in the masses and widths of the four states listed in Table~\ref{tab:Poles_ccc} with respect to the free parameter $\mu$.
Within a considerable range of the $\mu$ values, the poles of these four states are located below the thethresholds of corresponding channel $D\Xi_{cc}$, $D\Xi_{cc}^{*}$, $D^{*}\Xi_{cc}$ and $D^{*}\Xi_{cc}^{*}$, respectively.
As the value of $\mu$ increases, the masses decrease, while the widths first increase and then decrease. 
From the results of Fig. \ref{fig:ccc}, one can see that these systems are stably bound for a reasonable range of the $\mu$ values. 
Thus, our prediction for these bound systems are valuable for future experiments. 

As one can see from the results shown in Table \ref{tab:Poles_bbb}, the results for the $\Omega_{bbb}$-like sector are similar to those for the $\Omega_{ccc}$-like sector except for the mass differences, which have similar interactions as given in Table \ref{tab:Coefficients} derived from analogous dynamics with the flavor symmetry. 
The four candidates for the three-beauty molecular pentaquarks mostly couple to the channels $\bar{B}\Xi_{bb}$, $\bar{B}\Xi_{bb}^{*}$, $\bar{B}^{*}\Xi_{bb}$, and $\bar{B}^{*}\Xi_{bb}^{*}$, respectively, and also strongly couple to the channels $\bar{B}_{s}\Omega_{bb}$, $\bar{B}_{s}\Omega_{bb}^{*}$, $\bar{B}_{s}^{*}\Omega_{bb}$, and $\bar{B}_{s}^{*}\Omega_{bb}^{*}$, separately. 
They are all bound by about $44-47$ MeV with very small widths, all of which are less than $3$ MeV.
Among them, the state of $\bar{B}\Xi_{bb}^{*}$, $(15602.71 - 1.37i)$ MeV, can decay into the channels $\eta\Omega_{bbb}$ and $\eta^{'}\Omega_{bbb}$, and the one of $\bar{B}^{*}\Xi_{bb}^{*}$, $(15649.73 - 0.82i)$ MeV, into the channels $\omega\Omega_{bbb}$ and $\phi\Omega_{bbb}$.
Thus, these two predicted molecular pentaquark candidates can be searched in such corresponding decay channels in future experiments.

\begin{table}[htbp]
\centering
\renewcommand\tabcolsep{2.0mm}
\renewcommand{\arraystretch}{1.50}
\caption{The poles (in MeV), couplings $|g_{i}|$ and compositeness $|1-Z|_{i}$ for each channel in the $\Omega_{bbb}$-like sector.}
\begin{tabular*}{86mm}{@{\extracolsep{\fill}}l|ccccc}
\toprule[1.00pt]
\toprule[1.00pt]
\multicolumn{6}{c}{\mbox{$\Omega_{bbb}$-like sector}}\\
\hline
&Channels&$\bar{B}\Xi_{bb}$&$\bar{B}_{s}\Omega_{bb}$&&\\
\hline
\multirow{3}{*}{\centering$0(\frac{1}{2}^{-})$}&\multicolumn{5}{l}{\mbox{$15574.29$ $(++)$}}\\
&$|g_{i}|$&$\textbf{1.39}$&$0.99$&&\\
&$|1-Z|_{i}$&$\textbf{0.58}$&$0.42$&&\\
\hline
&Channels&$\eta\Omega_{bbb}$&$\eta^{'}\Omega_{bbb}$&$\bar{B}\Xi_{bb}^{*}$&$\bar{B}_{s}\Omega_{bb}^{*}$\\
\hline
\multirow{3}{*}{\centering$0(\frac{3}{2}^{-})$}&\multicolumn{5}{l}{\mbox{$15602.71-1.37i$ $(--++)$}}\\
&$|g_{i}|$&$0.12$&$0.08$&$\textbf{1.32}$&$1.08$\\
&$|1-Z|_{i}$&$0.001$&$0.00$&$\textbf{0.52}$&$0.51$\\
\hline
&Channels&$\bar{B}^{*}\Xi_{bb}$&$\bar{B}_{s}^{*}\Omega_{bb}$&&\\
\hline
\multirow{3}{*}{\centering$0(\frac{1}{2}^{-},\frac{3}{2}^{-})$}&\multicolumn{5}{l}{\mbox{$15621.06$ $(++)$}}\\
&$|g_{i}|$&$\textbf{1.39}$&$0.99$&&\\
&$|1-Z|_{i}$&$\textbf{0.59}$&$0.41$&&\\
\hline
&Channels&$\omega\Omega_{bbb}$&$\phi\Omega_{bbb}$&$\bar{B}^{*}\Xi_{bb}^{*}$&$\bar{B}_{s}^{*}\Omega_{bb}^{*}$\\
\hline
\multirow{3}{*}{\centering$0(\frac{1}{2}^{-},\frac{3}{2}^{-},\frac{5}{2}^{-})$}&\multicolumn{5}{l}{\mbox{$15649.73-0.82i$ $(--++)$}}\\
&$|g_{i}|$&$0.04$&$0.12$&$\textbf{1.34}$&$1.06$\\
&$|1-Z|_{i}$&$0.00$&$0.002$&$\textbf{0.54}$&$0.47$\\
\bottomrule[1.00pt]
\bottomrule[1.00pt]
\end{tabular*}
\label{tab:Poles_bbb}
\end{table}

Note that there are some predictions from the other work. 
For the $\Omega_{ccc}$-like sector, a $D\Xi_{cc}$ bound state with the mass about $4310-4330$ MeV was predicted in Ref. \cite{Hofmann:2005sw}, which was at least about $150$ MeV below the threshold.
In Ref. \cite{Romanets:2012hm}, a similar state with the mass at $4358.2$ MeV was found, which mainly coupled to the $D\Xi_{cc}$ channel. 
These two predictions are more bound than what we obtain for the $D\Xi_{cc}$ state with a mass $5446.97$ MeV. 
Using the one-boson-exchange model, Ref. \cite{Chen:2017jjn} found two candidates for the molecules of $D\Xi_{cc}$ and $D^{*}\Xi_{cc}$ with the binding energies of $0.22$ MeV and $18.71$ MeV, respectively.
Compared to their results, the states we obtained are more bound, compared with our binding energy about $40$ MeV.
For the $\Omega_{bbb}$-like sector, a $\bar{B}\Xi_{bb}^{*}$ bound state at $15212.04$ MeV with quantum number $J^{P}=\frac{3}{2}^{-}$ was predicted in Ref. \cite{Dias:2019klk}.
It is below the threshold of the $\bar{B}\Xi_{bb}^{*}$ channel around $300$ MeV, which is much bigger than our binding energy of about $47$ MeV.
The main reason is that we use different scheme to regularize the loop functions compared to theirs. 
In addition, there may be some differences between the particle masses used in Ref. \cite{Dias:2019klk} and ours, which can be seen from the corresponding thresholds. 

\begin{table*}[htbp]
\centering
\renewcommand\tabcolsep{2.0mm}
\renewcommand{\arraystretch}{1.50}
\caption{The poles (in MeV), couplings $|g_{i}|$ and compositeness $|1-Z|_{i}$ for each channel in the $\Omega_{bcc}$-like sector.}
\begin{tabular*}{178mm}{@{\extracolsep{\fill}}l|ccccccccc}
\toprule[1.00pt]
\toprule[1.00pt]
\multicolumn{10}{c}{\mbox{$\Omega_{bcc}$-like sector}}\\
\hline
$0(\frac{1}{2}^{-})$&Channels&$\eta\Omega_{bcc}$&$\eta^{'}\Omega_{bcc}$&$D\Xi_{bc}$&$D\Xi_{bc}^{'}$&$\bar{B}\Xi_{cc}$&$D_{s}\Omega_{bc}$&$D_{s}\Omega_{bc}^{'}$&$\bar{B}_{s}\Omega_{cc}$\\
\hline
$8764.63$&$|g_{i}|$&$0.00$&$0.00$&$\textbf{1.57}$&$0.00$&$0.00$&$1.52$&$0.00$&$0.00$\\
$(-+++++++)$&$|1-Z|_{i}$&$0.00$&$0.00$&$\textbf{0.78}$&$0.00$&$0.00$&$0.22$&$0.00$&$0.00$\\
$8768.18-12.71i$&$|g_{i}|$&$0.59$&$0.06$&$0.00$&$\textbf{2.00}$&$0.48$&$0.00$&$0.94$&$0.25$\\
$(-+++++++)$&$|1-Z|_{i}$&$0.03$&$0.00$&$0.00$&$\textbf{0.88}$&$0.02$&$0.00$&$0.08$&$0.003$\\
$8875.63-2.27i$&$|g_{i}|$&$0.21$&$0.06$&$0.00$&$0.10$&$\textbf{2.15}$&$0.00$&$0.51$&$1.29$\\
$(-+--++++)$&$|1-Z|_{i}$&$0.004$&$0.001$&$0.00$&$0.002$&$\textbf{0.88}$&$0.00$&$0.03$&$0.10$\\
$8983.70-42.36i$&$|g_{i}|$&$0.82$&$0.35$&$0.00$&$0.28$&$0.22$&$0.00$&$\textbf{1.83}$&$1.41$\\
$(------++)$&$|1-Z|_{i}$&$0.05$&$0.03$&$0.00$&$0.01$&$0.01$&$0.00$&$\textbf{0.72}$&$0.17$\\
\hline
$0(\frac{3}{2}^{-})$&Channels&$\eta\Omega_{bcc}^{*}$&$\eta^{'}\Omega_{bcc}^{*}$&$D\Xi_{bc}^{*}$&$\bar{B}\Xi_{cc}^{*}$&$D_{s}\Omega_{bc}^{*}$&$\bar{B}_{s}\Omega_{cc}^{*}$&&\\
\hline
$8793.07-12.20i$&$|g_{i}|$&$0.57$&$0.06$&$\textbf{2.00}$&$0.43$&$0.95$&$0.27$&&\\
$(-+++++)$&$|1-Z|_{i}$&$0.03$&$0.00$&$\textbf{0.88}$&$0.01$&$0.08$&$0.003$&&\\
$8928.31-3.65i$&$|g_{i}|$&$0.27$&$0.08$&$0.06$&$\textbf{2.15}$&$0.62$&$1.23$&&\\
$(-+-+++)$&$|1-Z|_{i}$&$0.01$&$0.001$&$0.001$&$\textbf{0.88}$&$0.06$&$0.09$&&\\
$9009.05-43.28i$&$|g_{i}|$&$0.82$&$0.35$&$0.31$&$0.31$&$\textbf{1.80}$&$1.46$&&\\
$(----++)$&$|1-Z|_{i}$&$0.05$&$0.03$&$0.01$&$0.01$&$\textbf{0.72}$&$0.16$&&\\
\hline
$0(\frac{1}{2}^{-},\frac{3}{2}^{-})$&Channels&$\omega\Omega_{bcc}$&$\phi\Omega_{bcc}$&$D^{*}\Xi_{bc}$&$D^{*}\Xi_{bc}^{'}$&$\bar{B}^{*}\Xi_{cc}$&$D_{s}^{*}\Omega_{bc}$&$D_{s}^{*}\Omega_{bc}^{'}$&$\bar{B}_{s}^{*}\Omega_{cc}$\\
\hline
$8909.83$&$|g_{i}|$&$0.00$&$0.00$&$1.49$&$0.00$&$0.00$&$\textbf{1.51}$&$0.00$&$0.00$\\
$(-+++++++)$&$|1-Z|_{i}$&$0.00$&$0.00$&$\textbf{0.78}$&$0.00$&$0.00$&$0.22$&$0.00$&$0.00$\\
$8915.24-10.17i$&$|g_{i}|$&$0.58$&$0.58$&$0.00$&$\textbf{1.65}$&$1.23$&$0.00$&$0.92$&$0.61$\\
$(-+++++++)$&$|1-Z|_{i}$&$0.04$&$0.04$&$0.00$&$\textbf{0.65}$&$0.25$&$0.00$&$0.07$&$0.02$\\
$8927.25-0.73i$&$|g_{i}|$&$0.15$&$0.13$&$0.00$&$0.88$&$\textbf{1.76}$&$0.00$&$0.68$&$1.32$\\
$(-+++++++)$&$|1-Z|_{i}$&$0.003$&$0.003$&$0.00$&$0.23$&$\textbf{0.69}$&$0.00$&$0.04$&$0.10$\\
$9115.00-33.41i$&$|g_{i}|$&$0.48$&$0.78$&$0.00$&$0.20$&$0.21$&$0.00$&$\textbf{1.76}$&$1.18$\\
$(-----+++)$&$|1-Z|_{i}$&$0.02$&$0.10$&$0.00$&$0.01$&$0.003$&$0.00$&$\textbf{0.65}$&$0.22$\\
\hline
$0(\frac{1}{2}^{-},\frac{3}{2}^{-},\frac{5}{2}^{-})$&Channels&$\omega\Omega_{bcc}^{*}$&$\phi\Omega_{bcc}^{*}$&$D^{*}\Xi_{bc}^{*}$&$\bar{B}^{*}\Xi_{cc}^{*}$&$D_{s}^{*}\Omega_{bc}^{*}$&$\bar{B}_{s}^{*}\Omega_{cc}^{*}$&&\\
\hline
$8942.49-10.64i$&$|g_{i}|$&$0.57$&$0.55$&$\textbf{1.79}$&$0.60$&$1.08$&$0.08$&&\\
$(-+++++)$&$|1-Z|_{i}$&$0.04$&$0.04$&$\textbf{0.79}$&$0.04$&$0.11$&$0.00$&&\\
$8977.58-0.27i$&$|g_{i}|$&$0.09$&$0.11$&$0.24$&$\textbf{2.00}$&$0.40$&$1.32$&&\\
$(-+++++)$&$|1-Z|_{i}$&$0.001$&$0.002$&$0.05$&$\textbf{0.84}$&$0.02$&$0.10$&&\\
$9141.48-35.55i$&$|g_{i}|$&$0.49$&$0.78$&$0.23$&$0.13$&$\textbf{1.75}$&$1.21$&&\\
$(----++)$&$|1-Z|_{i}$&$0.02$&$0.09$&$0.01$&$0.002$&$\textbf{0.68}$&$0.18$&&\\
\bottomrule[1.00pt]
\bottomrule[1.00pt]
\end{tabular*}
\label{tab:Poles_bcc}
\end{table*}

In Table \ref{tab:Poles_bcc}, we show the results for the $\Omega_{bcc}$-like sector.
For the case $PB(\frac{1}{2}^{+})$ with $J^{P}=\frac{1}{2}^{-}$, we find four bound states with the poles $8764.63$ MeV, $(8768.18 - 12.71i)$ MeV, $(8875.63 - 2.27i)$ MeV, and $(8983.70 - 42.36i)$ MeV.
These bound states mostly couple to the channels $D\Xi_{bc}$, $D\Xi_{bc}^{'}$, $\bar{B}\Xi_{cc}$, and $D_{s}\Omega_{bc}^{'}$, respectively.
Their binding energies are about $25$ MeV, $47$ MeV, $26$ MeV, and $32$ MeV, separately, which are small compared to their masses located at so high energy region.

For the case $PB(\frac{3}{2}^{+})$ with $J^{P}=\frac{3}{2}^{-}$, we obtain three states at $(8793.07-12.20i)$ MeV, $(8928.31-3.65i)$ MeV, and $(9009.05-43.28i)$ MeV.
They locate at about $25-47$ MeV below the thresholds of the most relevant channels $D\Xi_{bc}^{*}$, $\bar{B}\Xi_{cc}^{*}$, and $D_{s}\Omega_{bc}^{*}$, respectively. 
The width of the last state is relatively large, because several channels are open and can decay into.

For the case $VB(\frac{1}{2}^{+})$ degenerate in $J^{P}=\frac{1}{2}^{-}$, $\frac{3}{2}^{-}$, we also obtain four states, where the most relevant channels are the $D^{*}\Xi_{bc}$ for the one $8909.83$ MeV, the $D^{*}\Xi_{bc}^{'}$ for the state $(8915.24-10.17i)$ MeV, the $\bar{B}^{*}\Xi_{cc}$ for the pole $(8927.25-0.73i)$ MeV, and the $D_{s}^{*}\Omega_{bc}^{'}$ for the last one $(9115.00-33.41i)$ MeV, accordingly. 
Their binding energies are about $21$ MeV, $41$ MeV, $19$ MeV, and $44$ MeV, respectively.
Note that, for the state $D^{*}\Xi_{bc}$ with the mass $8909.83$ MeV, 
its coupling to the channel $D^{*}\Xi_{bc}$ is $g=1.49$, which is not the biggest one and a bit smaller than the one to the channel $D_{s}^{*}\Omega_{bc}$ of $g=1.51$.
Its mass $8909.83$ MeV is about $213$ MeV below the threshold of the $D_{s}^{*}\Omega_{bc}$ channel, where the binding energy for the $D_{s}^{*}\Omega_{bc}$ channel is too large with respect to the other ones. 
But, for the results of the compositenesses, the biggest one is 0.78 for the channel $D^{*}\Xi_{bc}$. 
In view of the results for the couplings and the compositenesses, the pole $8909.83$ MeV should be mainly contributed from the channel $D^{*}\Xi_{bc}$. 
Therefore, we make a further checking and investigate the interaction of the single channel, where we find that one pole of $D^{*}\Xi_{bc}$ is bound by $65$ MeV, and the other one of $D_{s}^{*}\Omega_{bc}$ is bound by $41$ MeV.
This test result indicates that the pole $8909.83$ MeV found in the coupled channel interaction is dominant by the channel $D^{*}\Xi_{bc}$. 
Furthermore, if we take the free parameter $\mu=1000$ MeV, the mass of this state becomes $8867.35$ MeV, and its couplings to the two channels are reversed, which are $g=1.94$ for the channel $D^{*}\Xi_{bc}$, and $g=1.79$ for the channel $D_{s}^{*}\Omega_{bc}$. 
Thus, in this case, one can easily see that the pole $8867.35$ MeV mostly couples to the channel $D^{*}\Xi_{bc}$ with the binding energy about $63$ MeV, which also confirm our conclusion above. 
Indeed, the channel $D^{*}\Xi_{bc}$ couples strongly to the $D_{s}^{*}\Omega_{bc}$, which leads to the large coupling $g=1.79$ for the pole  $8867.35$ MeV and can be seen from the non-diagonal coefficient $\sqrt{2}$ between them in Table~\ref{tab:Coefficients}. 
Note that, from the results of the coupled channel interaction as shown in Table~\ref{tab:Poles_bcc}, the pole of the channel $D_{s}^{*}\Omega_{bc}$ is missing, which can be expected from its attractive potential as given in Table~\ref{tab:Coefficients}. 
In fact, the channel $D_{s}^{*}\Omega_{bc}$ is loosely bound with the binding energy about $41$ MeV in the single channel interaction obtained above, which is smaller than the one $65$ MeV for the channel $D^{*}\Xi_{bc}$. 
Therefore, taking the parameter $\mu=800$ MeV for all the results at present, its pole has moved to the threshold and no stable pole to be found, which is a similar situation for the other systems with the states $\Omega_{cc}^{(*)}$, $\Omega_{bb}^{(*)}$, some of the systems with the $\Omega_{bc}^{(*)}$, and some of the cases with the $\Omega_{bc}^\prime$.

For the case $VB(\frac{3}{2}^{+})$ degenerate in $J^{P}=\frac{1}{2}^{-}$, $\frac{3}{2}^{-}$, $\frac{5}{2}^{-}$, we find three poles, where the most relevant channels are the $D^{*}\Xi_{bc}^{*}$ for the one $(8942.49-10.64i)$ MeV, the $\bar{B}^{*}\Xi_{cc}^{*}$ for the case $(8977.58-0.27i)$ MeV, and the $D_{s}^{*}\Omega_{bc}^{*}$ for the last one $(9141.48-35.55i)$ MeV.
Their binding energies are in the range of $22$ MeV to $39$ MeV.

\begin{table*}[htbp]
\centering
\renewcommand\tabcolsep{2.0mm}
\renewcommand{\arraystretch}{1.50}
\caption{The poles (in MeV), couplings $|g_{i}|$ and compositeness $|1-Z|_{i}$ for each channel in the $\Omega_{bbc}$-like sector.}
\begin{tabular*}{178mm}{@{\extracolsep{\fill}}l|ccccccccc}
\toprule[1.00pt]
\toprule[1.00pt]
\multicolumn{10}{c}{\mbox{$\Omega_{bbc}$-like sector}}\\
\hline
$0(\frac{1}{2}^{-})$&Channels&$\eta\Omega_{bbc}$&$\eta^{'}\Omega_{bbc}$&$D\Xi_{bb}$&$\bar{B}\Xi_{bc}$&$\bar{B}\Xi_{bc}^{'}$&$D_{s}\Omega_{bb}$&$\bar{B}_{s}\Omega_{bc}$&$\bar{B}_{s}\Omega_{bc}^{'}$\\
\hline
$12134.09-0.01i$&$|g_{i}|$&$0.01$&$0.14$&$\textbf{1.75}$&$0.01$&$0.01$&$1.28$&$0.004$&$0.003$\\
$(-+++++++)$&$|1-Z|_{i}$&$0.00$&$0.01$&$\textbf{0.62}$&$0.00$&$0.00$&$0.36$&$0.00$&$0.00$\\
$12189.43-7.63i$&$|g_{i}|$&$0.44$&$0.13$&$0.43$&$\textbf{1.59}$&$0.37$&$0.49$&$1.05$&$0.05$\\
$(--++++++)$&$|1-Z|_{i}$&$0.02$&$0.005$&$0.08$&$\textbf{1.10}$&$0.04$&$0.13$&$0.11$&$0.00$\\
$12217.65-1.30i$&$|g_{i}|$&$0.11$&$0.03$&$0.10$&$0.08$&$\textbf{1.32}$&$0.14$&$0.15$&$0.93$\\
$(----+-++)$&$|1-Z|_{i}$&$0.001$&$0.00$&$0.01$&$0.002$&$\textbf{0.91}$&$0.01$&$0.003$&$0.09$\\
\hline
$0(\frac{3}{2}^{-})$&Channels&$\eta\Omega_{bbc}^{*}$&$\eta^{'}\Omega_{bbc}^{*}$&$D\Xi_{bb}^{*}$&$\bar{B}\Xi_{bc}^{*}$&$D_{s}\Omega_{bb}^{*}$&$\bar{B}_{s}\Omega_{bc}^{*}$&&\\
\hline
$12163.34$&$|g_{i}|$&$0.01$&$0.14$&$\textbf{1.74}$&$0.01$&$1.29$&$0.01$&&\\
$(-+++++)$&$|1-Z|_{i}$&$0.00$&$0.01$&$\textbf{0.61}$&$0.00$&$0.37$&$0.00$&&\\
$12243.89-5.08i$&$|g_{i}|$&$0.23$&$0.07$&$0.21$&$\textbf{1.27}$&$0.29$&$0.98$&&\\
$(---+-+)$&$|1-Z|_{i}$&$0.004$&$0.001$&$0.03$&$\textbf{0.86}$&$0.04$&$0.10$&&\\
\hline
$0(\frac{1}{2}^{-},\frac{3}{2}^{-})$&Channels&$\omega\Omega_{bbc}$&$\phi\Omega_{bbc}$&$D^{*}\Xi_{bb}$&$\bar{B}^{*}\Xi_{bc}$&$\bar{B}^{*}\Xi_{bc}^{'}$&$D_{s}^{*}\Omega_{bb}$&$\bar{B}_{s}^{*}\Omega_{bc}$&$\bar{B}_{s}^{*}\Omega_{bc}^{'}$\\
\hline
$12233.59-1.54i$&$|g_{i}|$&$0.16$&$0.18$&$0.15$&$\textbf{1.42}$&$0.13$&$0.18$&$0.87$&$0.01$\\
$(--++++++)$&$|1-Z|_{i}$&$0.002$&$0.01$&$0.004$&$\textbf{0.91}$&$0.004$&$0.01$&$0.08$&$0.00$\\
$12262.25-0.63i$&$|g_{i}|$&$0.08$&$0.10$&$0.10$&$0.06$&$\textbf{1.34}$&$0.10$&$0.12$&$0.89$\\
$(--+-++++)$&$|1-Z|_{i}$&$0.001$&$0.003$&$0.002$&$0.002$&$\textbf{0.92}$&$0.002$&$0.002$&$0.08$\\
$12278.80-0.66i$&$|g_{i}|$&$0.09$&$0.09$&$\textbf{1.72}$&$0.01$&$0.01$&$1.27$&$0.01$&$0.01$\\
$(--+--+++)$&$|1-Z|_{i}$&$0.001$&$0.002$&$\textbf{0.63}$&$0.00$&$0.00$&$0.36$&$0.00$&$0.00$\\
\hline
$0(\frac{1}{2}^{-},\frac{3}{2}^{-},\frac{5}{2}^{-})$&Channels&$\omega\Omega_{bbc}^{*}$&$\phi\Omega_{bbc}^{*}$&$D^{*}\Xi_{bb}^{*}$&$\bar{B}^{*}\Xi_{bc}^{*}$&$D_{s}^{*}\Omega_{bb}^{*}$&$\bar{B}_{s}^{*}\Omega_{bc}^{*}$&&\\
\hline
$12283.84-2.78i$&$|g_{i}|$&$0.19$&$0.23$&$0.20$&$\textbf{1.45}$&$0.24$&$0.83$&&\\
$(--++++)$&$|1-Z|_{i}$&$0.003$&$0.01$&$0.01$&$\textbf{0.92}$&$0.01$&$0.07$&&\\
$12308.10-0.68i$&$|g_{i}|$&$0.09$&$0.10$&$\textbf{1.72}$&$0.01$&$1.27$&$0.02$&&\\
$(--+-++)$&$|1-Z|_{i}$&$0.001$&$0.002$&$\textbf{0.62}$&$0.00$&$0.37$&$0.00$&&\\
\bottomrule[1.00pt]
\bottomrule[1.00pt]
\end{tabular*}
\label{tab:Poles_bbc}
\end{table*}

In Table \ref{tab:Poles_bbc}, we show the poles and their couplings, and the compositenesses for each channel in the $\Omega_{bbc}$-like sector.
The results are still obtained by taking $\mu=800$ MeV for the loop functions.
We find a total of ten bound states here.
For the case $PB(\frac{1}{2}^{+})$ with $J^{P}=\frac{1}{2}^{-}$, the three states are the possible candidates of the molecules $D\Xi_{bb}$, $\bar{B}\Xi_{bc}$, and $\bar{B}\Xi_{bc}^{'}$, respectively, in view of the results of the couplings and compositenesses. 
For the case $PB(\frac{3}{2}^{+})$ with $J^{P}=\frac{3}{2}^{-}$, the two states are the possible molecules $D\Xi_{bb}^{*}$ and $\bar{B}\Xi_{bc}^{*}$, respectively.
For the case $VB(\frac{1}{2}^{+})$ degenerate in $J^{P}=\frac{1}{2}^{-}$, $\frac{3}{2}^{-}$, three possible molecular states are found in the channels $\bar{B}^{*}\Xi_{bc}$, $\bar{B}^{*}\Xi_{bc}^{'}$, and $D^{*}\Xi_{bb}$, respectively.
For the case $VB(\frac{3}{2}^{+})$ degenerate in $J^{P}=\frac{1}{2}^{-}$, $\frac{3}{2}^{-}$, $\frac{5}{2}^{-}$, the two bound systems are the $\bar{B}^{*}\Xi_{bc}^{*}$ and $D^{*}\Xi_{bb}^{*}$ with respective poles.
Most of the binding energies are about $10$ MeV more or less, and the ones for the other four are about $70$ MeV, which are larger than those of the $\Omega_{bcc}$ sector.

Furthermore, we study the dependence of the numerical results on the free parameter of the regularization scale $\mu$ in the loop functions.
We check with taking $\mu=q_{max}=650$ MeV, since three experimental $\Omega_{c}$ states were well reproduced in Refs. \cite{Debastiani:2017ewu,Debastiani:2018adr,Ikeno:2023uzz}.
The obtained poles with $\mu=650$ MeV are given in Table \ref{tab:Poles_650}.
The number of the molecular states dynamically generated from the coupled channel interactions is in agreement with the one with $\mu=800$ MeV, which means that these systems are stably bound. 
For the four states in the $\Omega_ {ccc}$-like sector, the binding energies decrease about $30$ MeV, while the change in the widths is minimal.
For the four bound systems in the $\Omega_ {bbb}$-like sector, the binding energies decrease about $13$ MeV.
For the ones in the $\Omega_ {bcc}$-like and $\Omega_ {bbc}$-like sectors, the binding energies decrease around $10-25$ MeV.
These results indicate that the numerical results we have before are not sensitive to the value of the parameter $\mu$ in the loop functions.
Thus, our results are stable and valuable for future experiments to look for more bound states in the high energy region. 

\begin{table*}[htbp]
\centering
\renewcommand\tabcolsep{1.5mm}
\renewcommand{\arraystretch}{1.50}
\caption{The poles (in MeV) for $\mu=650$ MeV.}
\begin{tabular*}{178mm}{@{\extracolsep{\fill}}lcccc}
\toprule[1.00pt]
\toprule[1.00pt]
$I(J^{P})$&$\Omega_{ccc}$-like&$\Omega_{bbb}$-like&$\Omega_{bcc}$-like&$\Omega_{bbc}$-like\\
\hline
\multirow{4}{*}{\centering$0(\frac{1}{2}^{-})$}&$5476.42$&$15586.94$&$8785.17$&$12159.85-0.07i$\\
&&&$8792.15-12.49i$&$12200.21-3.75i$\\
&&&$8892.04-1.58i$&$12225.56-0.67i$\\
&&&$9002.37-35.13i$&\\
\hline
\multirow{3}{*}{\centering$0(\frac{3}{2}^{-})$}&$5524.04-6.35i$&$15615.11-1.76i$&$8817.07-11.99i$&$12189.08-0.16i$\\
&&&$8944.62-2.43i$&$12251.39-3.06i$\\
&&&$9027.21-36.00i$&\\
\hline
\multirow{4}{*}{\centering$0(\frac{1}{2}^{-},\frac{3}{2}^{-})$}&$5619.74$&$15633.81$&$8928.30$&$12243.08-1.07i$\\
&&&$8939.37-8.41i$&$12270.50-0.37i$\\
&&&$8942.20-2.64i$&$12303.78-0.57i$\\
&&&$9132.93-28.28i$\footnote{Note that this pole was found on the $(------++)$ Riemann sheet, with the eighth channel not open, but the mass of this state is higher than its threshold, indicating that $\mu=650$ MeV may not be a reasonable parameter for this system.}&\\
\hline
\multirow{3}{*}{\centering$0(\frac{1}{2}^{-},\frac{3}{2}^{-},\frac{5}{2}^{-})$}&$5669.69-3.67i$&$15662.33-1.07i$&$8966.72-10.78i$&$12293.63-1.92i$\\
&&&$8992.69-0.56i$&$12333.04-0.59i$\\
&&&$9159.18-30.86i$&\\
\bottomrule[1.00pt]
\bottomrule[1.00pt]
\end{tabular*}
\label{tab:Poles_650}
\end{table*}

\section{Summary}\label{sec:Summary}

In recent years, the LHCb collaboration observed many candidates for the hidden-charm pentaquark, single-charm tetraquark, and double-charm tetraquark molecular states. 
Motivated by their findings in experiments, we try to search for the triple-heavy pentaquark systems in the present work. 

We systematically study the possible molecular pentaquark states with the triple-heavy flavor contents $cccq\bar{q}$ ($q=u,d,s$), $bbbq\bar{q}$, $bccq\bar{q}$, and $bbcq\bar{q}$.
The vector meson exchange mechanism for the meson-baryon interactions is taken into account with an extension of the local hidden gauge approach.
Then the $S$-wave scattering amplitudes are evaluated by solving the coupled channel Bethe-Salpeter equation. 
Note that in our theoretical model, there is only one free parameter in the loop functions, which is taken as $\mu=800$ MeV for all the results. 
Since the bound systems are only found in the isospin $I=0$ sector, 
we divide each sector into four blocks, $PB(\frac{1}{2}^{+})$ with quantum number $I(J^{P})=0(\frac{1}{2}^{-})$, $PB(\frac{3}{2}^{+})$ with $I(J^{P})=0(\frac{3}{2}^{-})$, $VB(\frac{1}{2}^{+})$ degenerate in $I(J^{P})=0(\frac{1}{2}^{-})$, $0(\frac{3}{2}^{-})$, and $VB(\frac{3}{2}^{+})$ degenerate in $I(J^{P})=0(\frac{1}{2}^{-})$, $0(\frac{3}{2}^{-})$ , and $0(\frac{5}{2}^{-})$.
In the $\Omega_{ccc}$-like sector, we obtain four candidates for the molecular states of the channels $D\Xi_{cc}$, $D\Xi_{cc}^{*}$, $D^{*}\Xi_{cc}$, and $D^{*}\Xi_{cc}^{*}$, respectively.
In the $\Omega_{bbb}$-like sector, we also obtain four bound systems of $\bar{B}\Xi_{bb}$, $\bar{B}\Xi_{bb}^{*}$, $\bar{B}^{*}\Xi_{bb}$, and $\bar{B}^{*}\Xi_{bb}^{*}$, separately.
In the $\Omega_{bcc}$-like sector, we find fourteen molecular states in the systems $D\Xi_{bc}$, $D\Xi_{bc}^{'}$, $\bar{B}\Xi_{cc}$,  $D_{s}\Omega_{bc}^{'}$, $D\Xi_{bc}^{*}$, $\bar{B}\Xi_{cc}^{*}$, $D_{s}\Omega_{bc}^{*}$, $D^{*}\Xi_{bc}$, $D^{*}\Xi_{bc}^{'}$, $\bar{B}^{*}\Xi_{cc}$, $D_{s}^{*}\Omega_{bc}^{'}$, $D^{*}\Xi_{bc}^{*}$, $\bar{B}^{*}\Xi_{cc}^{*}$, and $D_{s}^{*}\Omega_{bc}^{*}$, respectively.
In the $\Omega_{bbc}$-like sector, we get ten molecules of the channels $D\Xi_{bb}$, $\bar{B}\Xi_{bc}$, $\bar{B}\Xi_{bc}^{'}$, $D\Xi_{bb}^{*}$, $\bar{B}\Xi_{bc}^{*}$, $\bar{B}^{*}\Xi_{bc}$, $\bar{B}^{*}\Xi_{bc}^{'}$, $D^{*}\Xi_{bb}$, $\bar{B}^{*}\Xi_{bc}^{*}$, and $D^{*}\Xi_{bb}^{*}$, respectively.
The binding energies of these bound systems are in the order of $10-70$ MeV, and the widths of the most states are very narrow. 
Besides, there are some loose bound systems, such as the ones with the states $\Omega_{cc}^{(*)}$, $\Omega_{bb}^{(*)}$, some of the systems with the $\Omega_{bc}^{(*)}$, and some of the cases with the $\Omega_{bc}^\prime$, even though no stable pole is found in the present work.

In addition, we also investigate the uncertainty of our results with the change of the free parameter $\mu$. 
It is found that the bound systems obtained are stable. 
Thus, our predictions are helpful for future experiments to find new heavy states, and hope that future experiments can detect these predicted molecular states.

\section*{Acknowledgements}
 
We thank Professor Eulogio Oset for carefully reading the manuscript and providing valuable comments.
This work is supported by the National Natural Science Foundation of China (NSFC) under Grants No. 12335001, 12247101, 11965016 and 11705069, the National Key Research and Development Program of China under Contract No. 2020YFA0406400, the 111 Project under Grant No. B20063, the fundamental Research Funds for the Central Universities under Grant No. lzujbky-2022-sp02, and the project for top-notch innovative talents of Gansu province. This work is 
also partly supported by the Natural Science Foundation of Changsha under Grants No. kq2208257, the Natural Science Foundation of Hunan province under Grant No. 2023JJ30647, the Natural Science Foundation of Guangxi province under Grant No. 2023JJA110076, and the NSFC under Grant No. 12365019 (C.W.X.).

 \addcontentsline{toc}{section}{References}
 
\end{document}